\documentclass[a4paper,fleqn,usenatbib]{mnras}

\usepackage{newtxtext,newtxmath}

\usepackage[T1]{fontenc}
\usepackage{ae,aecompl}
\usepackage{gensymb}
\usepackage{booktabs,caption}
\usepackage[flushleft]{threeparttable}
\usepackage{longtable}
\usepackage{graphicx}	
\usepackage[graphicx]{realboxes}
\usepackage{amsmath}	
\usepackage{newtxtext,newtxmath}

\title[The superluminous SN~2018hti]{Close, bright and boxy: the superluminous SN 2018hti}

\author[A. Fiore et al.]{
A. Fiore,$^{1}$\thanks{E-mail: achillefiore@gmail.com}
S. Benetti,$^{1}$
M. Nicholl,$^{2,3}$
A. Reguitti,$^{{4,5,1}}$
E. Cappellaro,$^{1}$
S. Campana,$^{6}$
\newauthor
S. Bose,$^{7,8}$
E. Paraskeva,$^{9,10,11,{38}}$
E. Berger,$^{12}$
T. M. Bravo,$^{13}$
J. Burke,$^{14,15}$
Y.-Z. Cai,$^{16}$
\newauthor
T.-W. Chen,$^{17}$
P. Chen,$^{18}$
R. Ciolfi,$^{1,19}$
S. Dong,$^{18}$
S. Gomez,$^{20}$
M. Gromadzki,$^{21}$
\newauthor
C. P. Guti\'errez,$^{22,23}$
D. Hiramatsu,$^{14,15,{39,40}}$
G. Hosseinzadeh,$^{24}$
D. A. Howell,$^{14,15}$
\newauthor
A. Jerkstrand,$^{17}$
E. Kankare,$^{25}$
A. Kozyreva,$^{26}$
K. Maguire,$^{27}$
C. McCully,$^{14}$
\newauthor
P. Ochner,$^{1,28}$
C. Pellegrino,$^{14,15}$ 
G. Pignata,$^{4,5}$
R. S. Post,$^{29}$
{N. Elias-Rosa},$^{1,30}$
\newauthor
M. Shahbandeh,$^{31}$
S. Schuldt,$^{26,32}$
B. P. Thomas,$^{33}$
L. Tomasella,$^{1}$
J. Vink\'o,$^{33,34,35,36}$
\newauthor
C. Vogl,$^{26}$ 
J. C. Wheeler,$^{33}$
D. R. Young$^{37}$
\\
$^{1}$INAF – Osservatorio Astronomico di Padova, Vicolo dell’Osservatorio 5, I-35122 Padova, Italy\\
$^{2}$Birmingham Institute for Gravitational Wave Astronomy and School of Physics and Astronomy, University of Birmingham, Birmingham B15 2TT, UK \\
$^{3}$Institute for Astronomy, University of Edinburgh, Royal Observatory, Blackford Hill, EH9 3HJ, UK \\
$^{4}$Departamento de Ciencias F\'{i}sicas – Universidad Andres Bello, Avda. Rep\'{u}blica 252, Santiago, Chile\\
$^{5}$Millennium Institute of Astrophysics, Nuncio Monsenor S\'{o}tero Sanz 100, Providencia, Santiago, Chile \\
$^{6}$INAF – Osservatorio Astronomico di Brera, Via Bianchi 46, I-23807 Merate, Italy \\
$^{7}$Department of Astronomy, The Ohio State University, 140 W. 18th Avenue, Columbus, OH 43210, USA \\
$^{8}$Center for Cosmology and AstroParticle Physics (CCAPP), The Ohio State University, 191 W.Woodruff Avenue,
Columbus, OH 43210, USA \\
$^{9}$IAASARS, National Observatory of Athens, 15236, Penteli, Greece \\
$^{10}$Department of Astrophysics, Astronomy \& Mechanics, Faculty of Physics, National and Kapodistrian University of Athens, 15784 Athens, Greece \\
$^{11}$Nordic Optical Telescope, Apartado 474, E-38700 Santa Cruz de La Palma, Santa Cruz de Tenerife, Spain \\
$^{12}$Center for Astrophysics | Harvard \& Smithsonian, 60 Garden Street, Cambridge, MA 02138-1516, USA \\
$^{13}$School of Physics and Astronomy, University of Southampton, Southampton, Hampshire, SO17 1BJ, UK \\
$^{14}$Las Cumbres Observatory, 6740 Cortona Dr. Suite 102, Goleta, CA, 93117, USA \\
$^{15}$Department of Physics, University of California, Santa Barbara, Santa Barbara, CA, 93106, USA \\
$^{16}$Physics Department and Tsinghua Center for Astrophysics (THCA), Tsinghua University, Beijing, 100084, China \\
$^{17}$The Oskar Klein Centre, Department of Astronomy, Stockholm University, AlbaNova, SE-10691 Stockholm, Sweden \\
$^{18}$Kavli Institute for Astronomy and Astrophysics, Peking University, Yi He Huan Road 5, Hai Dian District, Beijing 100871, China \\
$^{19}$INFN -- Sezione di Padova, Via Francesco Marzolo 8, I-35131 Padova, Italy \\
$^{20}$Space Telescope Science Institute, 3700 San Martin Drive, Baltimore, MD 21218, USA \\
$^{21}$Astronomical Observatory, University of Warsaw, Al. Ujazdowskie 4, 00-478 Warszawa, Poland \\
$^{22}$Finnish Centre for Astronomy with ESO (FINCA), FI-20014 University of Turku, Finland \\
$^{23}$Tuorla Observatory, Department of Physics and Astronomy, FI-20014 University of Turku, Finland \\
$^{24}$Steward Observatory, University of Arizona, 933 North Cherry Avenue, Tucson, AZ 85721-0065, USA \\
$^{25}$Department of Physics and Astronomy, University of Turku, FI-20014 Turku, Finland\\
$^{26}$Max-Planck-Institut f\"ur Astrophysik, Karl-Schwarzschild-Str. 1, D-85748, Garching, Germany \\
$^{27}$School of Physics, Trinity College Dublin, The University of Dublin, Dublin 2, Ireland 
 \\
$^{28}$Dipartimento di Fisica e Astronomia G. Galilei , Universit\'{a} di Padova, Vicolo dell’Osservatorio 3, I-35122 Padova, Italy \\
$^{29}$Post Astronomy, Lexington, MA, USA \\
$^{30}$Institute of Space Sciences (ICE, CSIC), Campus UAB, Carrer de Can Magrans s/n, 08193 Barcelona, Spain \\
$^{31}$Department of Physics, Florida State University, 77 Chieftan Way, Tallahassee, FL 32306, USA \\
$^{32}$Technische Universit\"at M\"unchen, Physik Department, James-Franck Str. 
1, {85748} Garching, Germany \\
$^{33}$Department of Astronomy, University of Texas at Austin, 2515 Speedway, Stop C1400, Austin, Texas 78712-1205, USA \\
$^{34}$Konkoly Observatory,  CSFK, Konkoly-Thege M. \'ut 15-17, Budapest, 1121, Hungary \\
$^{35}$ELTE E\"otv\"os Lor\'and University, Institute of Physics, P\'azm\'any P\'eter s\'et\'any 1/A, Budapest, 1117 Hungary \\
$^{36}$Department of Optics \& Quantum Electronics, University of Szeged, D\'om t\'er 9, Szeged, 6720, Hungary \\
$^{37}$Astrophysics Research Centre, School of Mathematics and Physics, Queen's University Belfast, Belfast BT7 1NN, UK \\
$^{38}$Department of Physics and Astronomy, Aarhus University, NyMunkegade
120, DK-8000 Aarhus C, Denmark\\
$^{39}$Center for Astrophysics \textbar{} Harvard \& Smithsonian, 60 Garden Street, Cambridge, MA 02138-1516, USA \\
$^{40}$The NSF AI Institute for Artificial Intelligence and Fundamental Interactions\\\\\\
} 
\date{Accepted XXX. Received YYY; in original form ZZZ}
\pubyear{2015}
\begin{document}
\label{firstpage}
\pagerange{\pageref{firstpage}--\pageref{lastpage}}
\maketitle
\begin{abstract}
 SN~2018hti was a very nearby ($z=0.0614$) superluminous supernova with an exceedingly bright absolute magnitude of -21.7 mag in $r$-band at maximum. {The densely sampled pre-maximum light curves of SN~2018hti} show a slow luminosity evolution and constrain the rise time to $\sim$50 rest-frame days.
 We fitted synthetic light curves to the photometry to infer the physical parameters of the explosion of SN~2018hti for both the magnetar and the CSM-interaction scenarios. We conclude that one of two mechanisms could be powering the luminosity of SN~2018hti; interaction with $\sim10\,\mathrm{M}_\odot$ of circumstellar material or a magnetar with a magnetic field of $B_{\rm p}\sim1.3\times10^{13}$ G and initial period of $P_{\rm spin}\sim1.8$ ms. From the nebular spectrum modelling we infer that SN~2018hti likely results from the explosion of a $\sim40\,\mathrm{M}_\odot$ progenitor star.
\end{abstract}



\section{Introduction}
\label{sec:intro}
It is widely accepted that the explosion of massive stars \citep[$\gtrsim8\,\mathrm{M_\odot}$, e.~g.][]{smartt2009} is triggered by the gravitational collapse of their core{s}. This leads to a core-collapse supernova (SN) explosion, whose light curves (LCs) reach {an absolute} magnitude at maximum usually ranging between $-14$ and $-19$ mag \citep[e.g.][]{richardson2014,modjazetal2019} in optical bands. {These luminosities are suitably explained with the decay of $<0.1\,\mathrm{M_\odot}$ of $^{56}\mathrm{Ni}$ \citep[e.~g.][]{nadyozhin1994,muelleretal2017,anderson2019,prenticeetal2019} and with the thermal energy deposited in the progenitor's envelope during the gravitational collapse. The discoveries of superluminous supernovae (SLSNe) with an absolute magnitude even brighter than $-21$ mag \citep[e.g.][]{galyam2012,howell2017,galyam2019} challenge this standard supernova paradigm. In fact, $\gtrsim5\,\mathrm{M_\odot}$ of $^{56}\mathrm{Ni}$ would be required to account for these luminosities \citep[e.g.][]{kasenetal2011,dessartetal2012}}.

Apart from their exceptional brightness, SLSNe {are characterised by} their pre-maximum/maximum optical spectra, usually showing a hot ($\gtrsim15000$ K) continuum. {Similar} to the classical SNe \citep{minkowski1941}, {SLSNe} are subclassified as SLSNe~I and SLSNe~II depending on whether they are hydrogen deficient or hydrogen rich, respectively \citep{galyam2012}. In addition, SLSNe~IIn are characterized by the presence of multicomponent/narrow Balmer emission lines in {their} spectra and most likely fill the high luminosity tail of the luminosity function of SNe~IIn \citep{galyam2012}. 

SLSNe~I are usually discovered in metal-poor and star-forming host galaxies \citep{chenetal2013,lunnanetal2014,leloudasetal2015,perleyetal2015,chenetal2017a,schulzeetal2018}. {They are} recognized by the presence of prominent absorptions between 3000-5000 \AA{} in their pre-maximum/maximum optical spectra. This is an almost unique feature of SLSNe~I, usually identified as the contribution of O {\scriptsize II} transitions \citep[e.~g.][]{quimbyetal2011,mazzalietal2016,galyam2019b}, although this identification has been questioned \citep[e.g.][]{koenivesetal2020}. However, these features were observed also in the SN Ib SN~2008D \citep{soderbergetal2008}, {and} in the recently-proposed SN subclass of SNe Icn \citep[][]{galyametal2021,pastorelloetal2021} and in the type-II SN 2019hcc \citep{parragetal2021}. 15-20 days after maximum luminosity, the spectra of SLSNe I start to remarkably reproduce the behaviour of SNe Ic and SNe Ic broad lined (SNe Ic BL) at maximum luminosity \citep[e.g.][]{pastorelloetal2010}. Interestingly, recent discoveries of SLSNe I and SNe Ic {appear to} fill the luminosity gap between these two subclasses (such as the cases of the luminous SNe Ic SN~2012aa, SN~2019stc, \citealt{royetal2016}, \citealt{gomezetal2021}).  The physical explanation {linking} these SN subclasses is still {a} matter of investigations \citep[e.~g.][]{zouetal2018,blanchardetal2019,linetal2020b}. The photometric evolution of SLSNe~I is more heterogeneous: LCs of SLSNe~I {typically} evolve either smoothly \citep[e.~g. SN~2010gx, SN~2011ke, ][]{pastorelloetal2010,inserraetal2013} or they can show a complex behaviour with pre-/post-maximum bumps \citep[e.g. SN~2015bn, iPTF15esb, SN~2017gci, SN~2018don, ][{see also \citealt{hosseinzadehetal2021}}]{nicholletal2015,yanetal2015,angusetal2019,lunnanetal2020,fioreetal2021}. {Their} LCs evolve over a very broad range of {timescales; a diversity which} prompted the community to propose a slow/fast-evolving subclassification of SLSNe~I events, but the discovery of intermediate objects \citep[e.~g. Gaia16apd, SN~2017gci, ][]{kangasetal2017,yanetal2017a,nicholletal2017b,fioreetal2021,stevanceetal2021} and statistical studies \citep{nicholletal2015b,deciaetal2018,lunnanetal2018b,angusetal2019} point towards a {continuous} distribution between the two subcategories.

There is no general consensus about the engine powering SLSNe \citep[see][for a recent review]{moriyaetal2018}. Several scenarios {have been} proposed to explain the huge luminosities of SLSNe~I: (i) the magnetar scenario, which considers the contribution of the radiation-dominated wind inflated by a spinning down millisecond magnetar \citep[e.g.][]{kasenandbildsten2010,woosley2010,inserraetal2013,wangetal2015,chenetal2015,chenetal2017b,chenandsukhbold2016,nicholletal2017,margalitetal2018,vurmandmetzger2021}; (ii) the interaction of the SN ejecta with shells of circumstellar material \citep[CSM, e.g.][]{chevalierandfransson2003,chevalierandirwin2011,ginzburgandbalberg2012,chatzopoulosetal2013,nicholletal2014,smith2017,lunnanetal2018,lunnanetal2020,nicholletal2020} lost by the progenitor star prior to its explosion via stellar winds or alternatively via the pulsational-pair instability phenomenon \cite[e.g.][]{woosleyetal2007,woosley2017,renzoetal2020}; (iii) the pair-instability scenario, where $e^{+},e^{-}$ pair creation in a very massive star {\citep[with a He-core mass $64\,\mathrm{M}_\odot\lesssim\,M_{\rm He}\lesssim133\,\mathrm{M}_\odot$, e.g. ][]{hegerandwoosley2002}} induces the collapse of the star and triggers a thermonuclear runaway in the core, allowing for a massive production of $^{56}\mathrm{Ni}$. CSM interaction is usually invoked as the major power source for (SL)SNe~IIn \citep[as in the case of SN 2006gy, ][]{smithetal2007,smithetal2007b,agnolettoetal2009} as it provides a suitable explanation for the narrow/multicomponent features usually seen in their spectra. 
SLSNe~I LCs can be reasonably well explained by CSM models too \citep[e.~g.][]{chevalierandfransson2003,chevalierandirwin2011,ginzburgandbalberg2012,chatzopoulosetal2013}, although the SLSNe I spectra lack for strong interaction signatures. However, it has been shown \citep{chevalierandirwin2011,moriyaandtominaga2012,smithetal2015,andrewsandsmith2018,bhirombhakdietal2019} that a buried CSM interaction might suppress these features under {specific} conditions, e.~g. if the SN progenitor star is surrounded by a CSM disk \citep{smith2017}. {The complexities in some SLSNe LCs are also indicative of CSM interaction}. While the {simplest explanation for these complexities is late-time} interaction with shells or clumps of CSM \citep[e.g.][]{moriyaetal2018}, \citet{metzgeretal2014} argue that they can be mimicked by the opacity variations {due} to wind-driven ionisation fronts of a millisecond magnetar.

In this work, we present and discuss the spectrophotometric observations of SLSN~I SN~2018hti, located at $\mathrm{RA}=03^\mathrm{h}\,40^\mathrm{m}\,53.76^\mathrm{s}$, $\mathrm{Dec}=+11\degree\,46'37.17''$. SN~2018hti was discovered on 2018 November 1 by the Asteroid Terrestrial-impact Last Alert System (ATLAS) project \citep{tonryetal2018,tonryetal2018b} and initially named ATLAS18yff. {It was then} classified on 2018 November 6 by \citet{burkeetal2018} as a SLSN I. Independent spectro-photometric data of SN~2018hti {are} already presented by \citet{linetal2020a} {and} imaging polarimetry data of SN~2018hti {are} presented by \citet{lee2019}. Here we present the photometric and spectroscopic data of SN~2018hti in Sec.~\ref{sec:photo} and Sec.~\ref{sec:spec}, respectively. We discuss in Sec. \ref{sec:disc} {the spectro-photometric data of SN~2018hti}: in particular, the metallicity measurements of its host galaxy (Sec.~\ref{sec:metal}), the blackbody temperature and radius evolutions (Sec.~\ref{sec:tempevol}), the photospheric velocity (Sec.~\ref{sec:vphot}), some photometric and spectroscopic comparisons of SN~2018hti with a selected sample of SLSNe~I (Sec.~\ref{sec:comps}) and {finally} the suitability of magnetar and of CSM-interaction scenarios for SN~2018hti (Sec.~\ref{sec:dataint}).

In the following, we will assume a flat Universe with $H_0=71\pm3\,\mathrm{km\,s^{-1}\,Mpc^{-1}}$, $\Omega_\Lambda=0.69$, $\Omega_\mathrm{M}=0.31$ {\citep[{taking} an average of $H_0$ among the estimates provided by][]{planck2016,kethanetal2021,riessetal2021}}. Hence, the redshift $z=0.0614$ (see Sect.~\ref{sec:spec}) measured with the narrow emission lines from the host galaxy corresponds to a luminosity distance $d_\mathrm{L}=271.2^{+12}_{-11}\,\mathrm{Mpc}$.

\section{Photometry}
\label{sec:photo}
\subsection{Observations and data reduction}
 We led the multi-band photometric follow up of SN~2018hti via several facilities. Ultraviolet ($uvw2,uvm2,uvw1$) and $U,B,V$ imaging was obtained with the {\textit{Neil Gehrels Swift} Observatory+UVOT \citep{gehrelsetal2004}}. Optical/near-infrared (NIR) $u,B,g,V,r,i,z,J,H,K_{\rm s}$ photometric follow-up was obtained via the NOT Unbiased Transient Survey\footnote{ \texttt{http://nuts2.sn.ie} .}  \citep[NUTS/NUTS2, ][]{mattilaetal2016,holmboetal2019} at the {2.56-}m Nordic Optical Telescope (NOT)+ALFOSC/NOTCam at the Roque de los Muchachos Observatory, La Palma (Spain), the 1.82m Copernico Telescope+AFOSC and Schmidt Telescopes at the Asiago Astrophysical Observatory (Italy){, the 1.2 meters telescope at the Fred Lawrence Whipple Observatory+KeplerCam {and} the 0.6/0.8-meter telescopes at the Post Observatory (CA, USA) and Post Observatory Mayhill (NM, USA)}. {We also include the} the Asteroid Terrestrial-impact Last Alert System (ATLAS)-{photometry, the early ZTF public photometry available via the \textsc{irsa}\footnote{\texttt{https://irsa.ipac.caltech.edu/}.} archive} and Las Cumbres Observatory (LCO)-network\footnote{\texttt{https://lco.global/}.}  $U,B,g,V,r,i$ photometry. {LCO data {\citep{brownetal2013}} come from the Global Supernova Project.} ATLAS- $o$ and $c$ magnitudes were converted to standard Sloan $g$- and $r$-filter following \citet[][cfr. equation 2 therein]{tonryetal2018} and \citet[][cfr. equation 6 and Tab. 6 therein]{tonryetal2012}. {As the colour transformations used within these equations are determined from a stellar spectral energy distribution (SED), the conversion tends to increase the uncertainty of the resulting magnitudes.} Also, at very early epochs the $g-r$ colour was estimated via a colour extrapolation since the coeval $g,r$-filter photometry {is} not available. {Lastly, we included the mid-infrared (MIR) photometry observed with the Wide-field Infrared Survey Explorer (WISE) NASA mission in the $W1$ and $W2$ wavelength bands.  }

{Photometry was} performed with the \textsc{ecsnoopy} package\footnote{{\textsc{ecsnoopy}} is a package for SN photometry using PSF fitting and/or template subtraction developed by E. Cappellaro. A package description can be found at \texttt{http://sngroup.oapd.inaf.it/ecsnoopy.html}.
} \citep{cappellaro2014} using the point spread function (PSF) fitting technique. A detailed description of the image-processing procedures can be found in \citet{fioreetal2021}. In particular, for SN~2018hti we removed the background contamination using the template-subtraction technique in the $u,U,B,g,V,r,i,z,W1,W2$-filter images. This {was} performed with \textsc{ecsnoopy} via \textsc{hotpants} \citep{becker2015}. Suitable deep template $u,U,B,V$-filter frames were obtained at the NOT via NUTS2 on 2020 February 25, namely 414 rest-frame days after maximum light {and we used PanSTARRS $g,r,i,z$ pre-explosion images as template frames}. We assumed that SN~2018hti already faded well below the detection limit and used these frames as templates. {The $W1,W2$ frames used as template frames for the WISE photometry were {obtained} by the WISE mission on 2018 August 18, 19 (MJD=58348.35, 58349.47), i. e. before the explosion of SN~2018hti. }For the NIR frames, {the background level was interpolated with a low-order polynomial} since no suitable deep template frame in $J,H,K_{\rm s}$ band was available. {$B,g,V,r,i,z$ magnitudes were calibrated having evaluated the photometric zero points and colour terms with a sequence of field stars from the Pan-STARRS \citep[Panoramic Survey Telescope and Rapid Response System,][]{chambersetal2016} catalogue. {The WISE magnitudes were calibrated with their instrumental zero-points}. Calibrated PanSTARRS magnitudes were converted to standard SDSS system following \citet[][see equation 6]{tonryetal2012}. $u$ magnitudes could not be calibrated with the SDSS survey {as} SN~2018hti was located outside of its sky coverage. Hence we calibrated the $u$ magnitudes of the local field stars {against $u$-band photometry of Sloan standards fields take on the same photometric night}. For $U,B,V$ images the calibration was done after {converting} the Pan-STARRS magnitudes to Sloan {magnitudes} as before, and then {from} Sloan magnitudes to Johnson system following \citet{chonisandgaskell2008}. NIR magnitudes were calibrated with a local sequence of stars from the Two-Micron All Sky Survey \citep{skrutskieetal2006}.} \textit{Swift}/UVOT $uvw2,uvm2,uvw1,U,B,V$-filter magnitudes were measured by stacking the layers of the individual observing segments with the task \textsc{uvotimsum}. We then measured the brightness using {a} 2 arcsec-radius aperture with the task \textsc{uvotsource} {task} in \textsc{heasoft} version 6.25 \citep{Heasoft2014a}. To calibrate the \textit{Swift}/UVOT magnitudes, we used the recently-updated version (November 2020) of the sensitivity corrections. We {also} analysed data from the {\it Swift} X-ray telescope {by first stacking all 29 UVOT exposures}. The total amount of observing time amounts to 52.3 ks. {No source was detected at the location of SN 2018hti}. The $3\,\sigma$ upper limit on the 0.3-10 keV count rate at the SN position is $6.6\times 10^{-4}$ counts s$^{-1}$.
Assuming a power-law X-ray spectrum and the Galactic column density of $1.6\times 10^{21}$ cm$^{-2}$ and the distance given in Sec. \ref{sec:intro}, 
we derive an upper limit on the 0.3-10 keV unabsorbed luminosity of $4\times10^{41}$ erg s$^{-1}$.
This {is} the maximum mean luminosity the SN could have had during the entire {\it Swift} campaign.
Under the hypothesis that the putative X-ray emission follows the UV emission, we restricted our analysis to a time interval centred on the 
UV peak time in a $\pm6$ d around maximum. We selected 5 observations for a total exposure time of 7.9 ks. 
The SN is still undetected with a $3\,\sigma$ upper limit on the 0.3-10 keV count rate of $4.7\times 10^{-3}$ counts s$^{-1}$,
corresponding to a 0.3-10 keV unabsorbed luminosity of $3\times10^{42}$ erg s$^{-1}$.

\begin{figure*}
    \centering
    \includegraphics[width=0.85\textwidth]{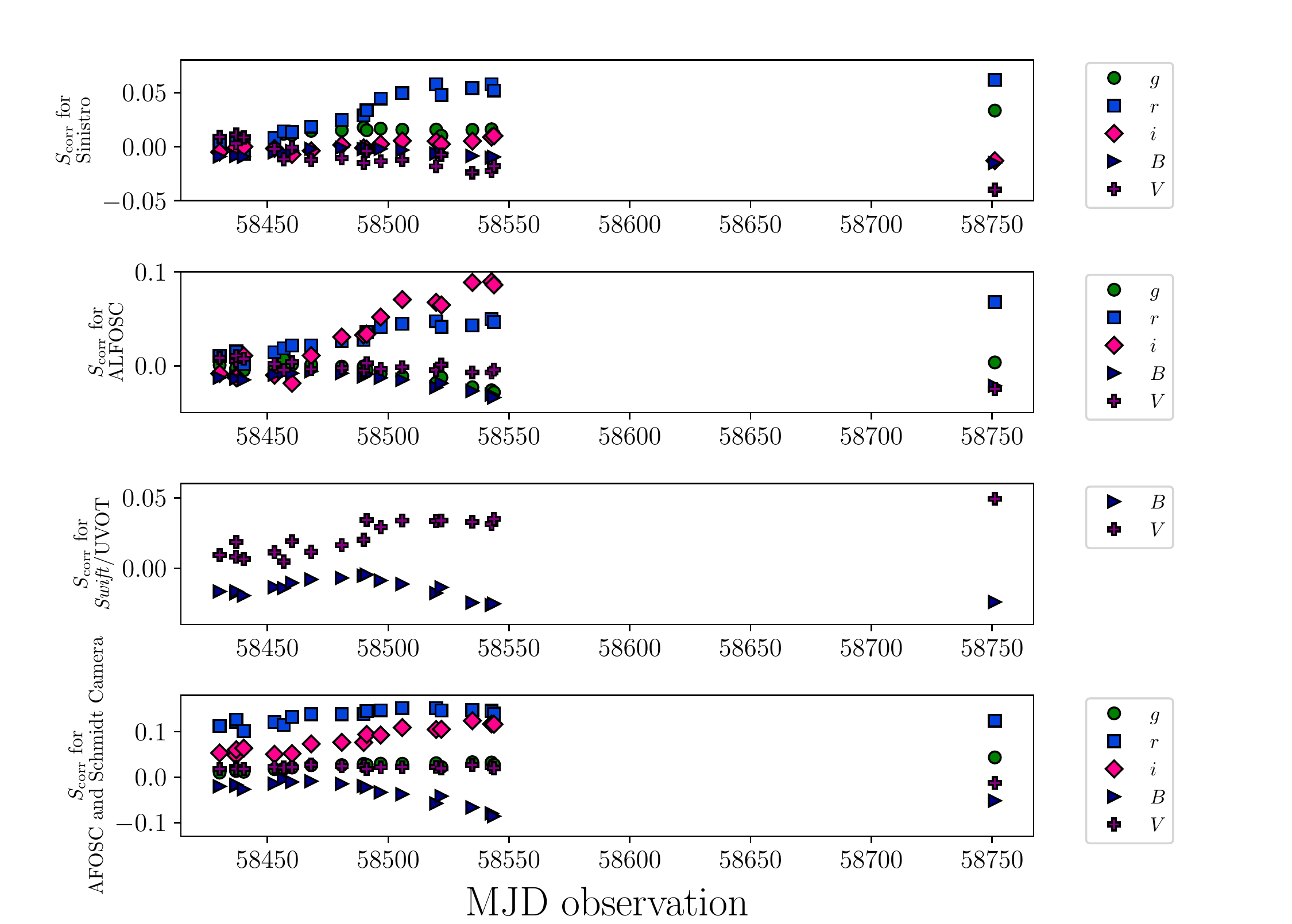}
    \caption{S-correction in $B,V,g,r,i$-filter passbands for different instruments (Sinistro, ALFOSC, \textit{Swift}/UVOT, AFOSC/Schmidt Camera, in descending order).}
    \label{fig:scorr}
\end{figure*}

Each instrument used for the observational follow-up has its own instrumental throughput. This difference introduces systematic errors {when magnitudes are obtained with multiple instruments}. To account for this effect, we compute the $B,V,g,r,i$ S-corrections \citep{stritzingeretal2002} for each instrumental configuration using the observed optical spectra of SN~2018hti \cite[similar to][]{pignataetal2004,eliasrosaetal2006,fioreetal2021} and propagated this {into} the calculation of the pseudo-bolometric LC. However, we noticed that this correction does not affect our analysis. 
The resulting S-correction for the $B,V,g,r,i$ filters and for each instrumental setup is shown in Fig.~\ref{fig:scorr}.
For the $u,U,z,J,H,K_{\rm s}$ filters (which are not covered by the observed {optical/NIR} spectra), we repeated the above procedure for a set of blackbody spectra {shifted} to the observer frame of SN~2018hti. The blackbody spectra have {temperatures} spanning a range of 8000-25000 K which broadly agrees with the best-fit blackbody temperatures of the spectra of SN~2018hti (see Sec.~\ref{sec:tempevol}). In this way, the S-corrections computed {for} the blackbody spectra provide an estimate of the S-correction outside the optical range. Unfortunately, this approach does not account for the presence of broad emission lines in the SN spectrum, which may alter the estimate of the S-correction. Synthetic-photometry measurements on the two available UV/NIR maximum/post-maximum spectra of other SLSNe I (Gaia16apd, \citealt{kangasetal2017}, and a NIR spectrum of the SN Ic BL SN 1998bw) show that the S-corrections computed on the spectra agree with those computed on their blackbody fit within $\sim0.05$ mag for the $z, J, H, K_{\rm s}$ filters. In the $u$ and $U$ bands, the blackbody approximation overcorrects the magnitudes with respect to the spectra of Gaia16apd (probably {due to} the line blanketing). This is true also for the NOTCam S-corrections calculated on the NIR IRTF+SpeX spectrum. {To carefully account} for this effect requires a denser and better-sampled spectroscopic follow-up outside the optical range, which at the moment has not been done for SLSNe. {We therefore opted for the conservative approach of propagating the maximum S-correction $\Delta S_{\rm corr}$ computed for the blackbody fits into the error of} the pseudo-bolometric LC calculation (see Sec.~\ref{sec:olc}).

The reduced $uvw2,uvm2,uvw1,u,U,B,g,V,r,i,z,J,H,K_\mathrm{s},$ $W1,W2$ magnitudes are reported in Tab.~\ref{tab:18hti_uvottab}, \ref{tab:18hti_ugriztab}, \ref{tab:18hti_bvtab}, \ref{tab:18hti_jhktab}, \ref{tab:18hti_wise}. The S-corrections $S_{\rm corr}$ and the $\Delta S_{\rm corr}$ values are listed in Tab.~\ref{tab:scorrasiago},\ref{tab:scorrlco},\ref{tab:scorrnot}, \ref{tab:scorrswift},\ref{tab:deltascorr}. 
\subsection{K-corrections}
We computed the $K$-corrections of SN~2018hti for the $B,V,g,r,i$-filter magnitudes performing synthetic photometry measurements onto the observed-frame ($m_{s,{\rm obs}}$) and rest-frame ($m_{s,{\rm rest}}$) optical spectra (see Sec.~\ref{sec:spec}). These were performed with the \textsc{pysynphot} \textsc{python} package\footnote{\texttt{https://pysynphot.readthedocs.io/} .}. For each filter and each spectrum, the $K$-corrections were computed as $K=m_{s,{\rm obs}}-m_{s,{\rm rest}}$ and are listed in Tab.~\ref{tab:18hti_kcorr}. {The measured $K$-corrections are linearly interpolated to the epochs of the imaging observation and subsequently subtracted from the magnitudes of SN 2018hti as measured on those images.}. $uvw2,uwm2,uvw1,u,U,z,J,H,K_{\rm s}$ $K$-corrections were estimated using the SED blackbody fits in place of the observed spectra.
\subsection{Observed and pseudo-bolometric light curves}
\label{sec:olc}
\begin{figure*}
\includegraphics[width=\textwidth]{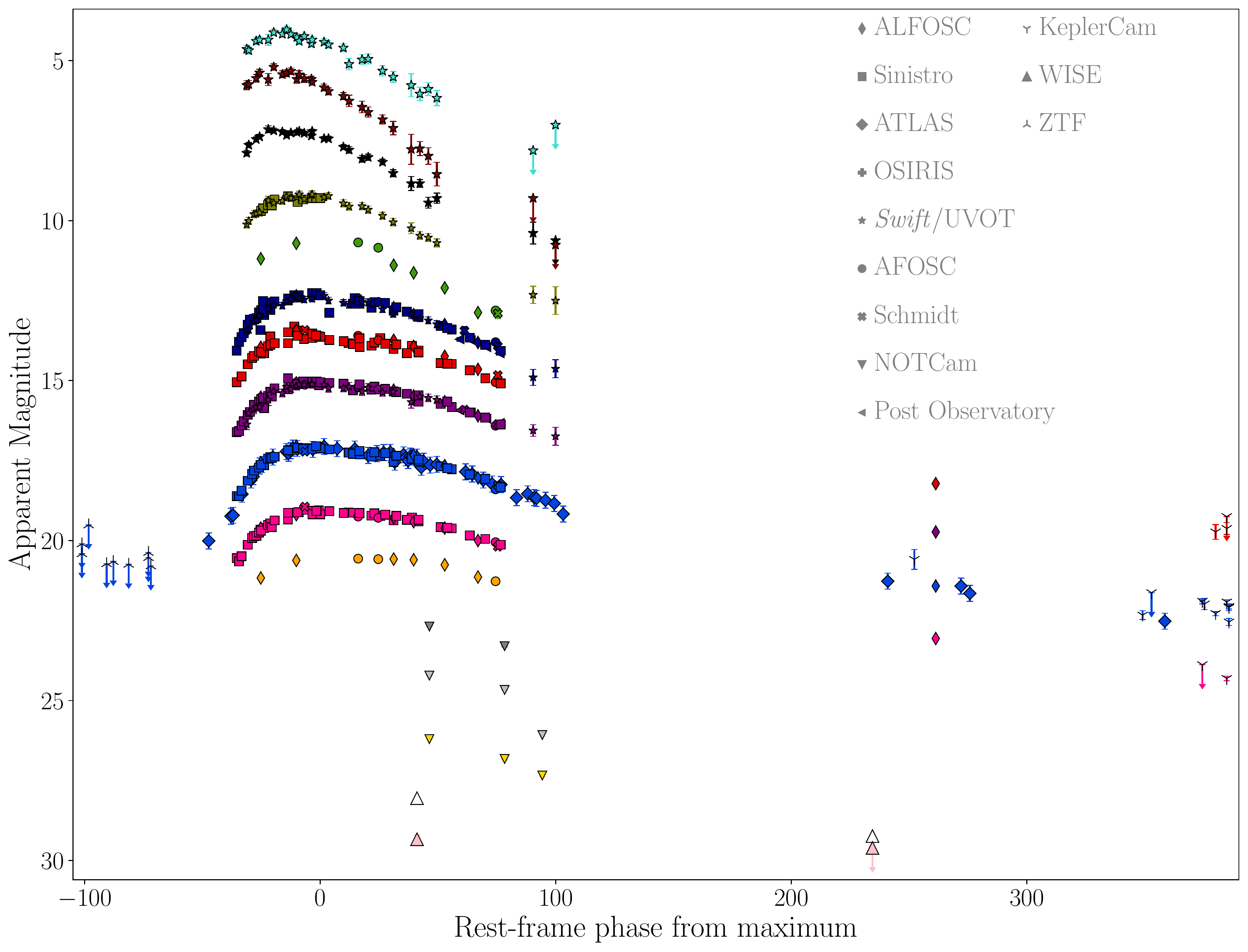}
\caption{$uvw2, uvm2, uvw1, U, u, B, g, V, r, i, z, J, H, K_{\rm s}, W1, W2$ observed LCs of SN~2018hti, respectively plotted in brown, cyan, black, dark green, dark blue, green, blue, red, purple, magenta, orange, silver, yellow, {white and pink}. Data obtained with different instruments are plotted with different markers, as labelled in the top-right corner. Magnitudes are plotted in ABmags.}
\label{fig:lcs}
\end{figure*}

The UV-optical-NIR observed LCs of SN~2018hti are shown in Fig.~\ref{fig:lcs} and are plotted against the phase corrected for time dilation. To estimate the maximum luminosity epoch, we fit a fourth order polynomial to the $r$-band LC {and infer} a magnitude at maximum $r_\mathrm{max}\simeq16.5\pm0.2$ mag at $\mathrm{MJD}=58\,464.5\pm4.0$ (in agreement with the maximum found by \citealt{linetal2020a}). Given a distance modulus $\mu=37.17\pm0.1\,\mathrm{mag}$ and a Galactic extinction\footnote{Obtained via the NASA/IPAC Extragalactic Database \texttt{https://ned.ipac.caltech.edu/extinction\_calculator} .} $A_V=1.280$ mag \citep{schlaflyetal2011}, assuming the extinction law $A_\mathrm{V}=3.1\times E(B-V)$ \citep{cardellietal1988,cardellietal1989}, the absolute peak $r$-filter magnitude is $M_r=-21.7\pm0.2$ mag. We assume no internal extinction from the host galaxy, supported by the absence of the interstellar \ion{Na}{i}D doublet and by the fact that the H$\alpha$/H$\beta$ ratio {remains} similar to the expected value for case-B recombination \citep{grovesetal2012}.

\begin{figure}
\includegraphics[width=0.5\textwidth]{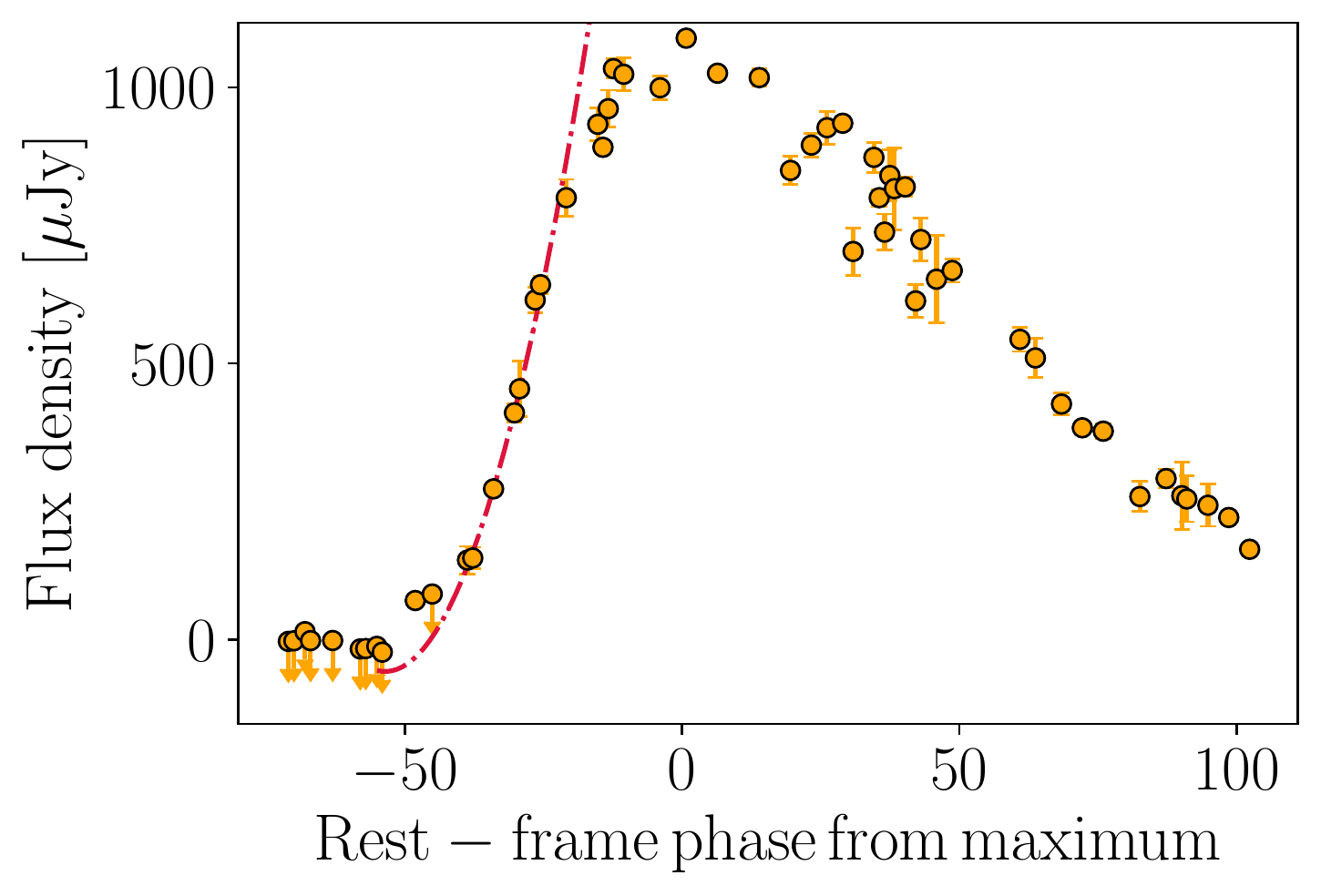}
\caption{ATLAS-$o$ LCs of SN~2018hti (yellow dots) in flux {space} (expressed in $\mu$Jy). The early ATLAS LC {is} fitted {with} a second-order polynomial (dashed dotted red line).}
\label{fig:lc_atlas}
\end{figure}

The earliest ATLAS detection was obtained on 2018 October 22 ($\mathrm{MJD}=58\,413.54$) and the last ATLAS detection limit was exactly three days before ($\mathrm{MJD}=58\,410.54$). If we assume that the latter {is} a genuine non detection, {this provides an} estimate of the explosion epoch {of} $\mathrm{MJD}=58\,412.04\pm1.5$. With this, the maximum-luminosity epoch implies a rest-frame rise time of $\tau_{\rm rise}=50\pm6$ days{, which is typical of the slow-evolving SLSNe~I \citep{inserra2019}}. Finally, after $\sim100$ days from maximum light, SN~2018hti {disappeared} behind the Sun.
For each filter, the LC evolution is characterized by a relatively slow rise to maximum and a {post-maximum decline rate of} $\sim1-2\times\tau_{\rm rise}$ in each filter. Moreover, the $r$-filter LC apparently shows a levelled off start at earliest phases, which {is hard to reconcile} with the overall trend of the LC. However, as noticeable in Fig.~\ref{fig:lcs}, the very early detections in $r$-filter were retrieved by the $o$-filter ATLAS magnitudes whose colour transformations is uncertain. {For this reason we also show} the observed ATLAS $o$-filter LC in Fig.~\ref{fig:lc_atlas} (top panel), which presents a similar slope change for the first point. To quantify the deviation of the first ATLAS detection from the early behaviour of the LC, we fit a parabola to the ATLAS flux density (expressed in $\mu$Jy) assuming a flux scaling $F\propto t^2$ \citep[e.~g.][]{riessetal1999,conleyetal2006}. Under this assumption, the first ATLAS point is $\sim$0.7 mags brighter than the predicted LC. 
{However, the early $r$- and ATLAS $o$-filter detection limits (see Figs.~ \ref{fig:lcs}, \ref{fig:lc_atlas}) exclude the occurrence of a pre-maximum bump \citep[as in ][]{leloudasetal2012,nicholletal2015,smithetal2016} up to $\sim53$ rest-frame days before the estimated explosion epoch.}
However, we {note} that the post-maximum epochs ATLAS $o$-filter data fluctuate within $\sim$0.25 mag in a timescale $\lesssim10$ days. Also, they do not apparently show up in other filters,
This allows for a 0.25 mag maximum uncertainty for ATLAS magnitudes, which is much less than the 0.7 mag deviation for the first point, making the flat start more credible.
Finally, the $K$-corrected and S-corrected $uvw2,uvm2,uvw1,u,U,B,g,V,r,i,z,J,H,K_{\rm s},W1,W2$ {host-template subtracted} photometry of SN~2018hti was {combined} to obtain the pseudo-bolometric LC displayed in Fig.~\ref{fig:bollc} (data are listed in Tab.~\ref{tab:18hti_blc}). This was computed by integrating the multiband photometry neglecting every flux contribution out of the integration boundaries. For SN~2018hti, the epochs of the $r$-band photometry {are} adopted as reference. The extinction corrected {combined} fluxes were finally converted {to} pseudo-bolometric luminosities by multiplying by $4\pi d_{\rm L}^2$. {Similar to} the multiband LCs, the pseudo-bolometric LC has a ratio $\tau_{\rm decline}/\tau_{\rm rise}\sim1.8$ \citep[similar to other SLSNe~I, see e.~g.][]{nicholletal2015b}.
\section{Spectroscopy}
\label{sec:spec}
\subsection{Observations and data reduction}
We collected a dense sample of spectra for SN~2018hti starting from $\mathrm{MJD}=58\,430.2$ (2018 November 8), which corresponds to 32 rest-frame days before maximum light. 

We led the spectroscopic follow up via the extended/advanced Public ESO Spectroscopic Survey for Transient Objects \citep{smarttetal2015} (ePESSTO/ePESSTO+), NUTS/NUTS2{, with the 1.82-m Copernico telescope at the Asiago astrophysical observatory, Italy, the 2.4m Hiltner Telescope+OSMOS (Ohio State Multi-Object Spectrograph) at the Michigan-Dartmouth-MIT Observatory and the Hobby Eberly Telescope (HET)+LRS2 (Low Resolution Spectrograph) at the McDonald observatory, Texas}. Moreover, we took a pre-maximum (-18 rest-frame days) NIR spectrum via the 3.0-m NASA Infrared Telescope Facility (IRTF)+SpeX \citep{rayneretal2003} and a nebular spectrum on 2019 September 24 (+269 rest-frame days) with the 10.4m Gran Telescopio CANARIAS (GTC)+OSIRIS \citep[Optical System for Imaging and low-Intermediate-Resolution Integrated Spectroscopy,][]{cepaetal2000} at Roque de los Muchachos Observatory. Additional FLOYDS spectra were obtained from FLOYDS on the Faulkes Telescope South (or North) as part of the Global Supernova Project. The instrumental setups and the resolution of the spectra presented in this work are listed in Tab.~\ref{tab:18hti_sfo}.

The {AFOSC, EFOSC2 and GTC} spectra were reduced with the standard \textsc{iraf} procedures. Firstly, the raw bidimensional spectroscopic frames were debiased, flatfielded and corrected for the cosmic-rays contribution with the Laplacian Cosmic Ray Identification package \citep[\textsc{lacosmic}, ][]{vandokkum2001}. Then the spectra were extracted along the spatial direction with the \textsc{iraf} task \textsc{apall} after having subtracted the background contribution, which was estimated {via} a low-order polynomial fit. The one-dimensional spectra were wavelength calibrated against HeAr (for the NTT+EFOSC2){, HeNe (for NOT+ALFOSC spectra)}, NeHgCd (for the 1.82m-Copernico+AFOSC spectra) and HgArNe (for GTC+OSIRIS spectrum) calibration arcs. Then the extracted one-dimensional spectra were flux calibrated via a set of spectrophotometric standard stars observed on the same night and with the same instrumental setup {as the science} observations. {Using the flux-calibrated standard star spectrum we were able to remove the contribution of the telluric absorption features}. Finally, the flux calibration was checked against coeval photometry. {The ALFOSC spectra were reduced with {\textsc{foscgui}}\footnote{{\textsc{foscgui} is a graphic user interface aimed at extracting SN spectroscopy and photometry obtained with FOSC-like instruments. It was developed by E. Cappellaro. A package description can be found at \texttt{http://sngroup.oapd.inaf.it/foscgui.html}.}}. The OSMOS spectrum was reduced with the \textsc{pyraf}-based \textsc{simspec}\footnote{\texttt{https://astro.subhashbose.com/simspec/ .}} pipeline. The LRS2 spectra were reduced with a dedicated \textsc{iraf}- and \textsc{python}-based pipeline \citep[as in ][ see Sec. 2.2.3]{yangetal2020}. {FLOYDS spectra were reduced using the \textsc{floydsspec} pipeline\footnote{\texttt{https://github.com/svalenti/FLOYDS\_pipeline/}\\\texttt{blob/master/bin/floydsspec/}}}. The IRTF+SpeX spectrum was reduced utilizing the \textsc{spextool} software package \citep{cushingetal2004}.}
\begin{figure}
\centering
\includegraphics[width=0.55\textwidth]{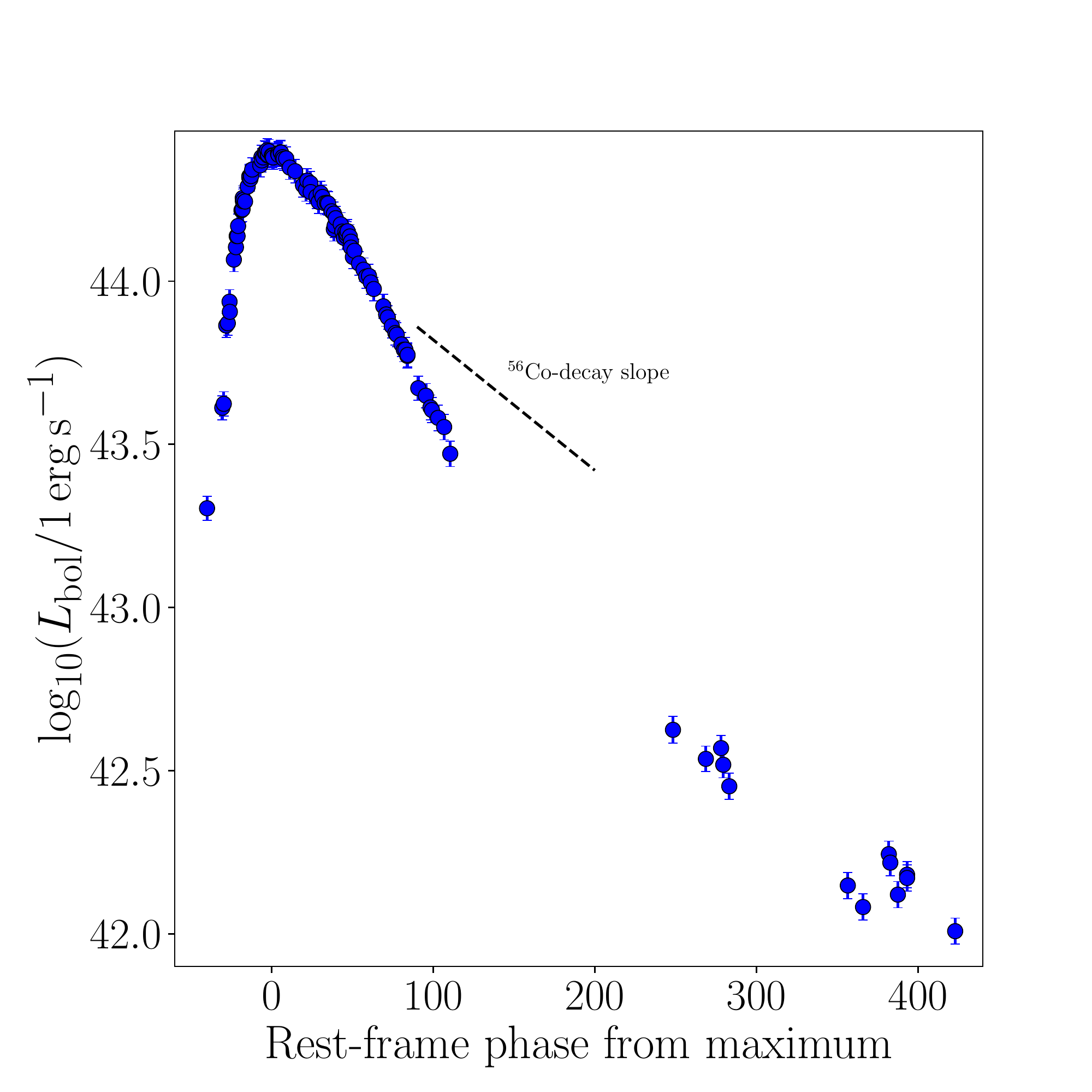}
\caption{Pseudo-bolometric LC of SN~2018hti (blue dots). The black-dashed line indicates the slope of $^{56}$Co-decay.}
\label{fig:bollc}
\end{figure}
\subsection{Spectral evolution and line identifications}
\begin{figure*}
\includegraphics[width=1\textwidth]{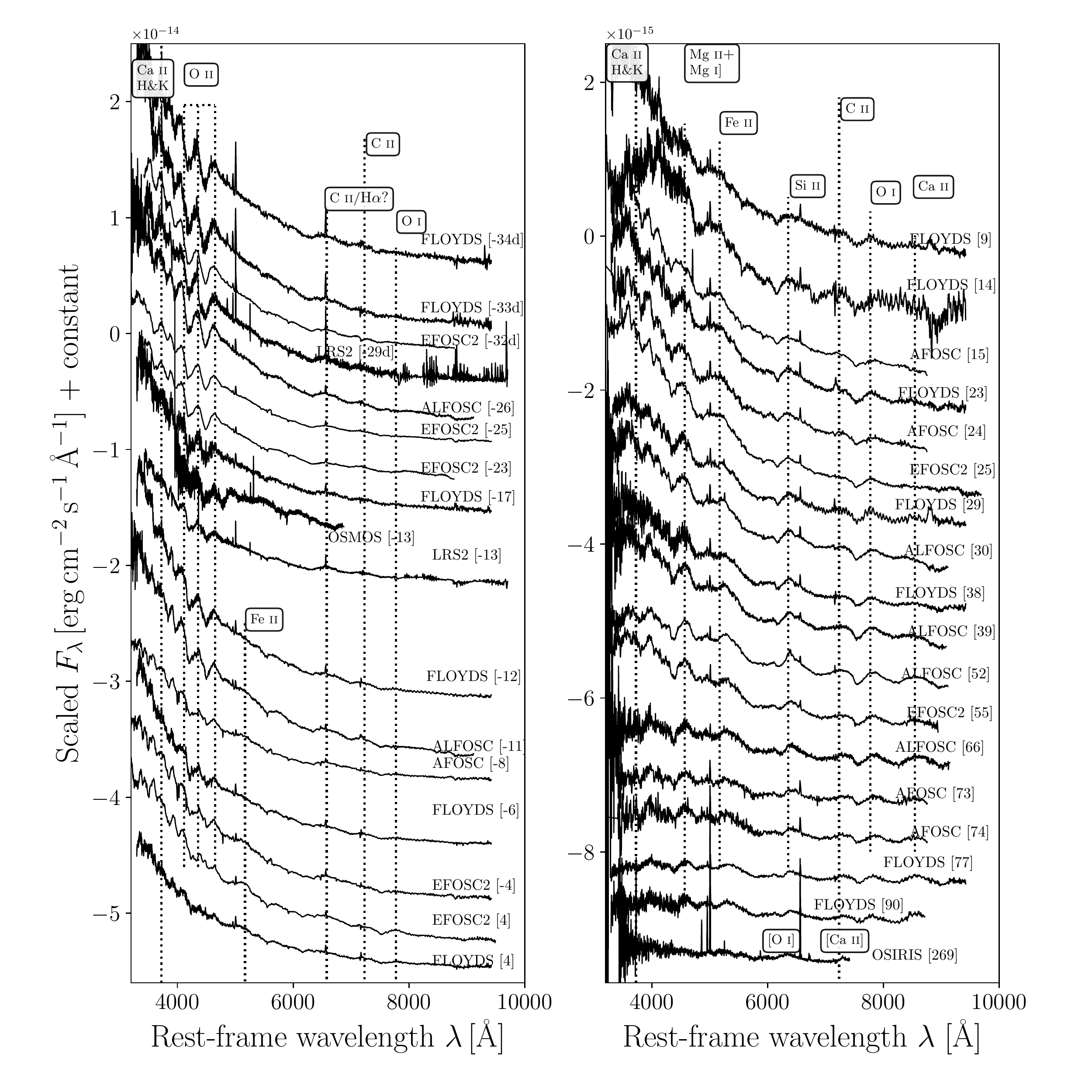}
\caption{Spectral evolution of SN~2018hti. Spectral line identifications are marked with vertical black dotted lines, and labelled on its right side with the corresponding ion. The rest-frame phase with respect to maximum luminosity is reported on the right side of each spectrum. The left panel shows the spectral evolution of SN~2018hti from -34 days to 4 days from maximum luminosity and the right panel shows the remaining spectra up to 269 days after maximum luminosity.}
\label{fig:spec}
\end{figure*}

The spectral evolution of SN~2018hti is shown in Fig.~\ref{fig:spec}. {Throughout} their evolution, the spectra of SN~2018hti show H$\beta$, H$\alpha$, [O {\scriptsize III}] $\lambda\,4959$, [O {\scriptsize III}] $\lambda\,5007$ narrow emission lines from the host galaxy, which we use to measure the redshift of the host galaxy and to estimate the metallicity at the site of SN~2018hti (see Sect.~\ref{sec:metal}). 

\begin{figure*}
    \centering
    \includegraphics[width=0.8\textwidth]{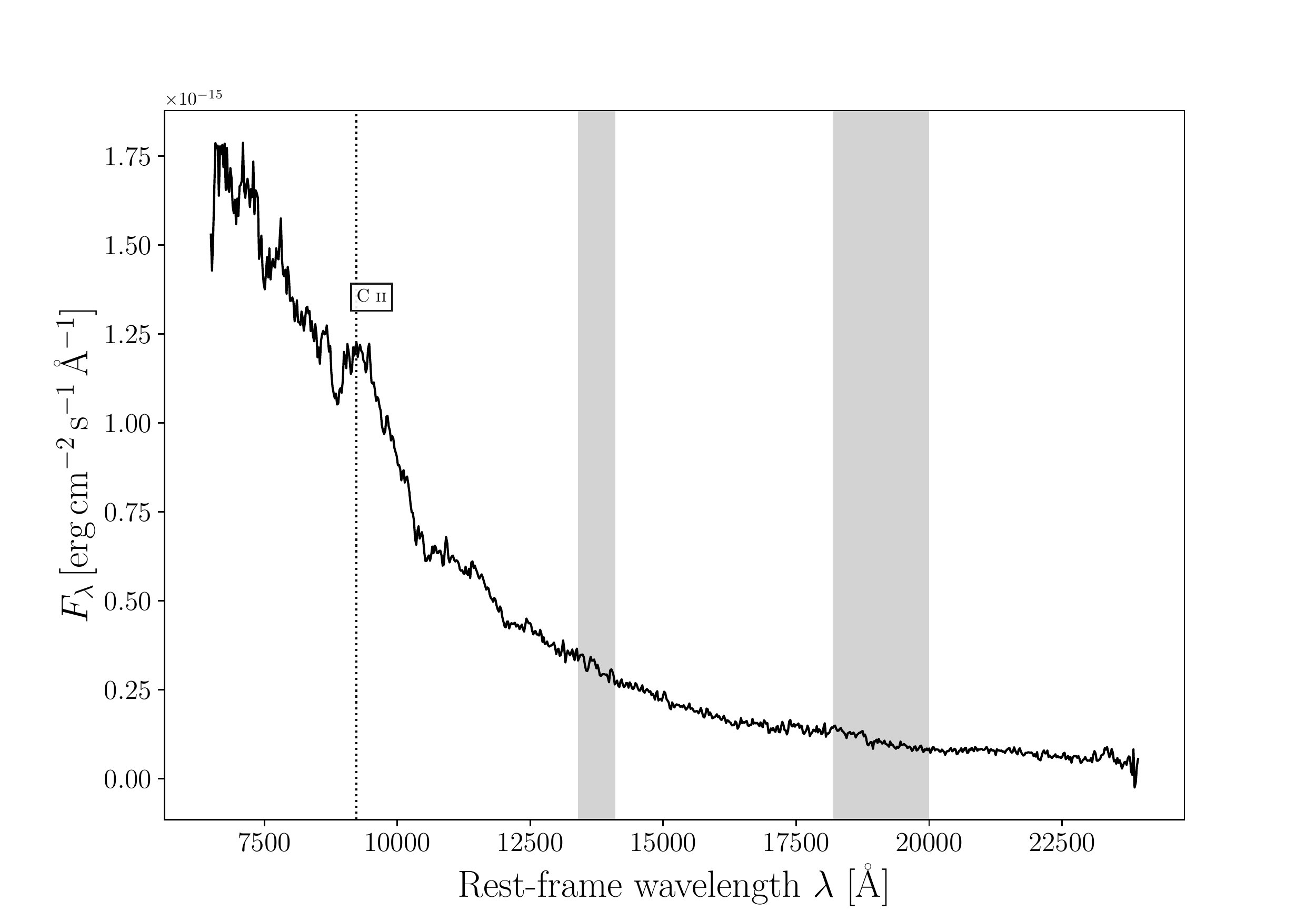}
    \caption{{The IRTF+SpeX spectrum of SN~2018hti (black solid line). The black dotted line marks the \ion{C}{ii} at $\lambda=9234$ \AA{} line identification and the shaded gray areas mark the spectral regions corrected for telluric absorptions.}}
    \label{fig:irtf}
\end{figure*}

The pre-maximum/maximum-light spectra of SN~2018hti present a very hot continuum reaching blackbody temperatures of $\sim18000-22000\,\mathrm{K}$. They show the W-shaped O {\scriptsize II} absorptions between 3500-5000 \AA{}. In addition, from the earliest spectrum at phase -34 days from the maximum light, a broad feature (FWHM$\sim15000-18000\,\mathrm{km\,s^{-1}}$) starts to rise in a nearly boxy fashion (see Sec.~\ref{sec:boxy}). We interpreted this feature as C {\scriptsize II} $\lambda\,6580$ \citep[as in][see also the discussion in Sec.~\ref{sec:boxy}]{nicholletal2014}. We also mention that the \textsc{tardis} modelling of a sample of more than 180 spectra of SLSNe I predicts {a} \ion{C}{ii} 6584.70 at a fraction above 50\% with a small contribution of \ion{Ne}{i} $\lambda\lambda$ 6404.02, 6508.83 (Paraskeva et al. in prep). This feature does not however reproduce a boxy profile.
{The early NIR spectrum of SN~2018hti (see Fig.~\ref{fig:irtf}) shows an almost featureless continuum with the exception of an emission at a rest-frame wavelength about $\simeq9200$ \AA{} which we interpreted as \ion{C}{ii} $\lambda\,9234$, similar to the cases of Gaia16apd \citep{yanetal2017a} and to SN~2015bn \citep{nicholletal2016}.}

In the spectrum taken 4 days after maximum light, the Fe {\scriptsize II} emission features are visible {in the blue regions, }while Mg {\scriptsize II} begins to be seen in the {15 day post-maximum spectrum}. On the same epoch, O {\scriptsize I} $\lambda\,7774$ {appears in} the red {end} of the spectrum.
About 15-30 days after maximum light, the spectrum of SN~2018hti smoothly enters the SN Ic/SNe Ic BL -like phase, similarly to many other SLSNe~I \citep[e.g.][]{pastorelloetal2010,inserraetal2013,galyam2019}. 
After 39 rest-frame days from maximum an emission shows up at $\sim6360$ \AA{}, which we interpreted as Si~{\scriptsize II} $\lambda$ 6355. 
{In the 52 day post-maximum spectrum} the Ca {\scriptsize II} NIR $\lambda\lambda\lambda\,8498,8542,8662$ triplet {becomes visible}. After SN~2018hti {reappeared from behind the sun}, we took the GTC+OSIRIS spectrum on 2019 September 24, 269 days {after} maximum light. This spectrum is not completely nebular {as} it 
displays some residual continuum, which could be however influenced by a residual contribution from the host galaxy \citep[see also][]{jerkstrandetal2017}. This phase was referred to as `pseudo-nebular' by \citet{nicholletal2019}.
\section{Discussion}
\label{sec:disc}
{Here} we discuss the data presented above. Where possible, we compare data of SN~2018hti with those of other SLSNe~I. To do this, we selected a sample of SLSNe~I which share some spectro-photometric properties with those of SN~2018hti. {We included LSQ14bdq \citep{nicholletal2015}, SN~2006oz \citep{leloudasetal2012} and DES14X3taz \citep{smithetal2016} since their $r$-filter LCs show a pre-maximum bump}. Moreover, SN~2015bn \citep{nicholletal2016} was prompted as the best-spectral match by \textsc{gelato} \citep{harutyunyanetal2008}. The SLSNe~I iPTF13ehe, iPTF15esb and iPTF16bad \citep{yanetal2015,yanetal2017b} were added to the comparison sample since they show H$\alpha$, although at later epochs with respect to maximum light. {Finally, we added also a late spectrum of Gaia16apd \citep{kangasetal2017} since few SLSNe~I spectra are available at pseudo-nebular/nebular phases.}
\subsection{Metallicity at the location of SN~2018hti}
\label{sec:metal}
As mentioned earlier, SLSNe~I usually explode in metal-poor, star-forming environments. Several metallicity diagnostics are calibrated {from} the emission lines emerging from the host-galaxy spectrum. In the case of SN~2018hti, we measured the flux emitted by the [O {\scriptsize II}] $\lambda\,3727$, H$\beta$, [O {\scriptsize III}] $\lambda\,4959$, [O {\scriptsize III}] $\lambda\,5007$, H$\alpha$ and [S {\scriptsize II}] $\lambda\,6717$ narrow emission lines emerging from the host galaxy in the nebular spectrum. To measure the flux emitted within the narrow emission lines, we extracted the host-galaxy spectrum close to the position of SN 2018hti {by placing} the aperture adjacent to the SN itself.

One of these indicators is referred to as $R_{23}$ \citep{pageletal1979}:
\begin{equation}
R_{23}=\frac{([\mbox{O \scriptsize II}] \lambda\,3727+[\mbox{O {\scriptsize III}}] \lambda\,4959,5007)}{\mathrm{H}\beta}\,.  
\end{equation}
Another indicator which is often used is the so-called N2O2 \citep{kewleyanddopita2002}:
\begin{equation}
\mathrm{N2O2}=\frac{[\mbox{N {\scriptsize II}}] \lambda\,6584}{[\mbox{O {\scriptsize II}}] \lambda\,3727}\,.
\end{equation}
For SN~2018hti we found $\log_{10}R_{23}=0.96$ and $\log_{10}(\mathrm{N2O2})=-1.26$. 
To measure the metallicity of the host galaxy at the site of SN~2018hti, we evaluated different metallicity estimators simultaneously thanks to the tool \textsc{pymcz}\footnote{The package can be found at \\ \texttt{https://github.com/nyusngroup/pyMCZ} .} presented by \citet{biancoetal2016}.
\begin{figure}
    \centering
    \includegraphics[width=0.5\textwidth]{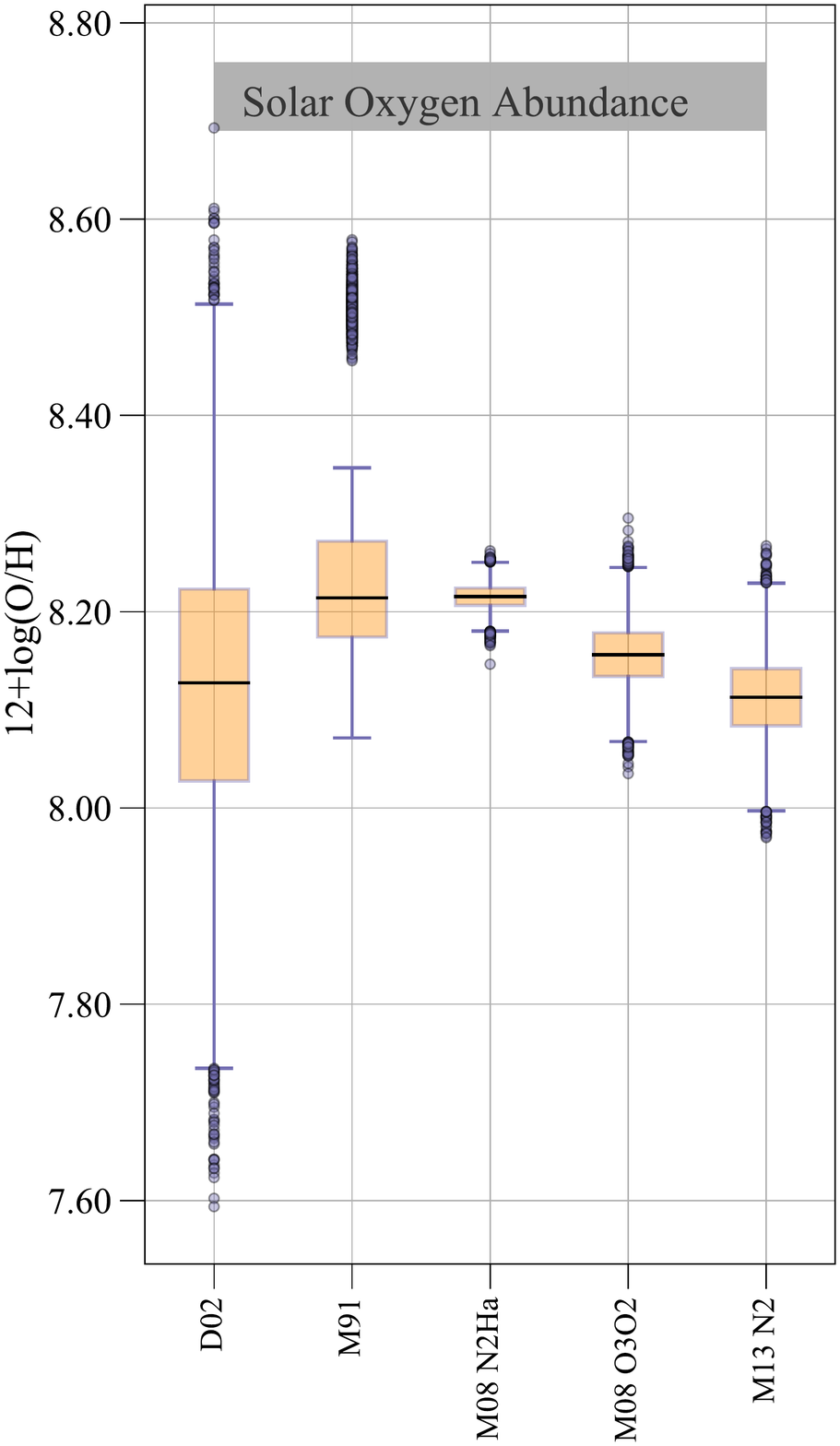}
    \caption{Boxplot obtained with the tool \textsc{pymcz}. The orange boxes cover the interquartile range (IQR) for each estimator, and the blue dots deviate from the first and third quartile more then $1.5\times\mathrm{IQR}$. The gray box broadly corresponds to the solar oxygen abundance.}
    \label{fig:pymcz}
\end{figure}
 \textsc{pymcz} randomly samples a Gaussian distribution whose mean and standard deviation are given by the flux measurements and their uncertainties, respectively. With this tool, it was possible to exploit the \textsc{d04} \citep{denicoloetal2002}, \textsc{m91} \citep{mcgaugh1991}, \textsc{m08\_n2ha}, \textsc{m08\_o3o2} \citep{maiolinoetal2008} and \textsc{m13\_n2} \citep{marinoetal2013} metallicity estimators. Other metallicity estimators {calculated by} \textsc{pymcz} are excluded from our analysis since they are not suitable for the case of SN~2018hti. In particular, \textsc{z94} \citep{zaritskyetal1994} is valid only for the upper branch of the $\log_{10}R_{23}$ scale\footnote{The upper-branch metallicity scale is defined by the condition $\log_{10}R_{23}<0.9$ \citep[e.g.][]{kewleyandellison2008}.}. Also, the \textsc{kd02} and \textsc{kk04} methods should only be used for  $\log_{10}(\mathrm{N2O2})>-1.2$. The results are shown in the boxplot in Fig.~\ref{fig:pymcz} and are summarized in Tab.~\ref{tab:pymcz}. As expected, the results point towards a metal-poor site with $12+\log_{10}(\mathrm{O/H})\approx8.17$, which corresponds to a metallicity  $Z\approx0.3\,Z_\odot$ \citep[assuming $12+\log_{10}(\mathrm{O/H})=8.69$ for the solar metallicity,][]{asplundetal2009}. This estimate nicely agrees with the results obtained by \citet{linetal2020a}.
\begin{table}
\centering
\begin{tabular}{ll}
\hline
estimator&$12+\log_{10}(\mathrm{O/H})$\\
\hline
\textsc{d02}&$8.128_{-0.149}^{+0.142}$\\
\rule{0pt}{3ex} 
\textsc{m91}&$8.214_{-0.055}^{+0.123}$\\
\rule{0pt}{3ex} 
\textsc{m08\_N2ha}&$8.215_{-0.013}^{+0.013}$\\
\rule{0pt}{3ex} 
\textsc{m08\_o3o2}&$8.156_{-0.033}^{+0.033}$\\
\rule{0pt}{3ex} 
\textsc{m13\_n2}&$8.113_{-0.042}^{+0.044}$\\
\hline
\end{tabular}
\caption{Metallicity estimators provided by the \textsc{pymcz} tool for the site of SN~2018hti (see also Fig.~\ref{fig:pymcz}).}
\label{tab:pymcz}
\end{table}
Moreover, we estimated the star formation rate (SFR) of the host galaxy of SN~2018hti based on the measurements of the flux emitted by the reddening corrected narrow H$\alpha$ using equation 2 of \citet{kennicutt1998}. The derived SFR is $\sim0.3\,\mathrm{M_\odot\,yr^{-1}}$, similar to the SFRs measured by \cite{chenetal2017a} for a sample of galaxies hosting SLSNe I and comparable to the SFR of the Large Magellanic Cloud \citep{harrisandzaritsky2009}. Finally, we compared the values of SFR and metallicity of SN~2018hti with those of the comparison sample (see Tab.~\ref{tab:metal}). Given the intrinsic uncertainty of these measurements, the selected SLSN-I sample {seem to} share similar environments, with the exception of SN~2015bn, which has a SFR about an order of magnitude lower than the others. However, as pointed out by \citet{nicholletal2015}, modelling the host-galaxy SED and estimating the median stellar mass and the age of the stellar population returns a higher SFR value of $0.55\pm0.18\,\mathrm{M_\odot\,yr^{-1}}$ for SN~2015bn. Also, the SFR value reported for LSQ14bdq is a SFR limit \citep[see also Sec. 4 in ][]{chenetal2017a}. 
\begin{table*}
\centering
\caption{Metallicities and SFRs of the SLSNe~I of the comparison sample as published in literature.}
\label{tab:metal}
\begin{tabular}{llllll}
\hline
&\textbf{SN~2018hti}&LSQ14bdq&SN~2015bn&SN~2006oz&DES14X3taz\\
\hline
$Z/Z_\odot$&\textbf{0.3}&-&0.2 \citep{nicholletal2016}&0.5 \citep{leloudasetal2012}&-\\
SFR $[\mathrm{M_\odot}\,\mathrm{yr}^{-1}]$&\textbf{0.3}&<0.05 \citep{chenetal2017a}&0.04 \citep{nicholletal2016}&0.17 \citep{leloudasetal2012}&0.16 \citep{smithetal2016}\\
\hline
\end{tabular}
\end{table*}
\subsection{Blackbody temperature and photospheric radius}
\label{sec:tempevol}
\begin{figure*}
    \centering
    \includegraphics[width=0.8\textwidth]{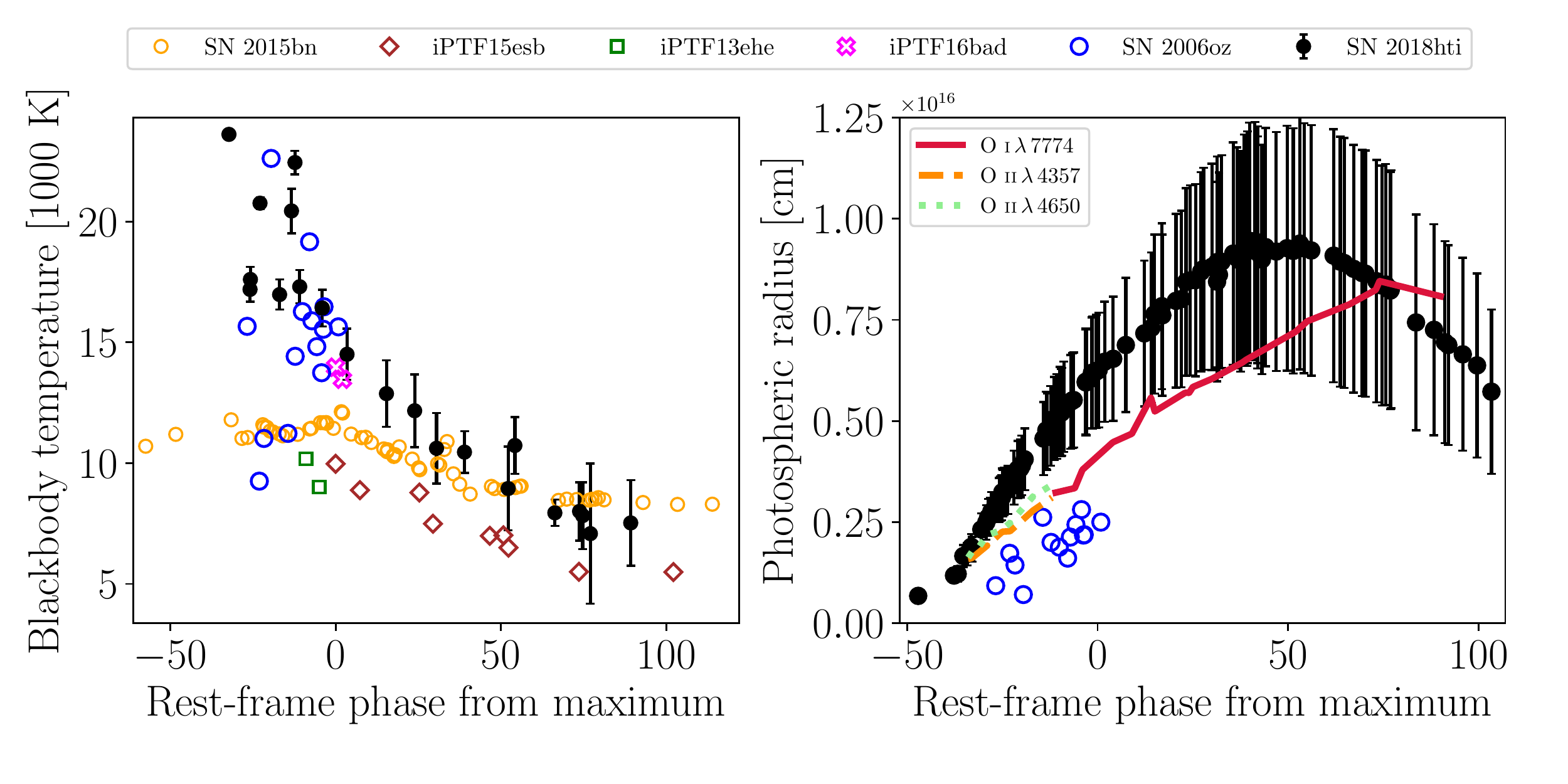}
    \caption{Left panel: the blackbody temperatures of SN~2018hti (black filled dots). The blackbody temperature evolution of SN~2015bn \citep[orange empty dots, ][]{nicholletal2016}, SN~2006oz \citep[blue empty dots, ][]{leloudasetal2012},  iPTF13ehe \citep[green empty squares, ][]{yanetal2015,yanetal2017b}, iPTF15esb \citep[brown empty diamonds, ][]{yanetal2017b} and iPTF16bad \citep[magenta empty crosses, ][]{yanetal2017b} are also shown for comparison. Right panel: the evolution of the photospheric radius of SN~2018hti (black dots) compared with SN~2006oz. For comparison we also plot the radius obtained from the photospheric-velocity measurements performed on the spectra for the \ion{O}{I} (red solid line) and the \ion{O}{II} absorptions ($\lambda$ 4357, orange dashed line, $\lambda$ 4650, green dotted line).}
    \label{fig:bbtemp}
\end{figure*}

We obtained the time evolution of the blackbody temperatures by fitting a blackbody curve to the spectra. This allow us to avoid the contribution of the spectral lines in the fitting procedure by excluding the line-contaminated regions from the fit domain. The comparison of the temperature evolution of SN~2018hti with SN~2015bn \citep{nicholletal2016}, SN~2006oz \citep{leloudasetal2012}, iPTF13ehe \citep{yanetal2015,yanetal2017b}, iPTF15esb \citep{yanetal2017b}, iPTF16bad \citep{yanetal2017b} is shown in Fig.~\ref{fig:bbtemp} (left panel). The data of SN~2006oz are relatively dispersed, but are useful for an order-of-magnitude comparison. The temperature evolution of SN~2018hti is essentially monotonic and is very similar to the case of SN~2006oz, the steepest of the sample. SN~2018hti and SN~2006oz appear to have the hottest photospheres among the SLSNe~I sample. However, the scarcer sampling of the blackbody temperatures of iPTF13ehe and iPTF16bad do not allow to properly compare them with SN~2018hti. In particular, both SN~2015bn and iPTF15esb reach a `temperature floor' \citep{nicholletal2017} of $5000-8000$ K after $\sim50-80$ days after maximum luminosity (Fig.~\ref{fig:bbtemp}) similar to the sample analysed by \citet{inserraetal2013} \citep[see also][]{nicholletal2017}. In the case of SN~2018hti it is unclear whether or not the temperature evolution actually settles on a plateau at that phase.

We also {determine} the evolution of the photospheric radius (Fig.~\ref{fig:bbtemp}, right panel) $R_{\rm ph}$ using the Stefan-Boltzmann law, where we used the pseudo-bolometric luminosities shown in Fig.~\ref{fig:bollc}. To {compare the more sparsely sampled} spectroscopic epochs {with those of} the pseudo-bolometric luminosities, we fit the blackbody temperatures with a second-order polynomial. The photospheric radius of SN~2018hti monotonically grows to a maximum value of $\sim9\times10^{15}$ cm in ~100 days, which is about $\sim50$ days later the maximum bolometric luminosity. It then recedes at a rate of about $7\times10^{13}\,\mathrm{cm^2\,\mathrm{day}^{-1}}$, which is similar to the average growth rate. Overall, the photospheric-radius evolution is consistent with the expansion radius determined from expansion velocity derived from some spectral lines (see Sec.~\ref{sec:vphot}) except for the time interval between -26 and 35 rest-frame days from maximum (see Fig.~\ref{fig:bbtemp}, right panel). However, given the huge uncertainties, this should be considered only as an order-of-magnitude comparison. In the same figure we also show the photospheric-radius evolution of SN~2006oz. The photosphere of SN~2006oz seems less extended than that of SN~2018hti. In fact, given that both of them have a comparable photospheric temperature, SN~2006oz is about $\sim${0.5} mag fainter than SN~2018hti (see Fig.~\ref{fig:lsq14bdq}). 
\subsection{Photospheric velocity}
\label{sec:vphot}
\begin{figure}
    \centering
    \includegraphics[width=0.5\textwidth]{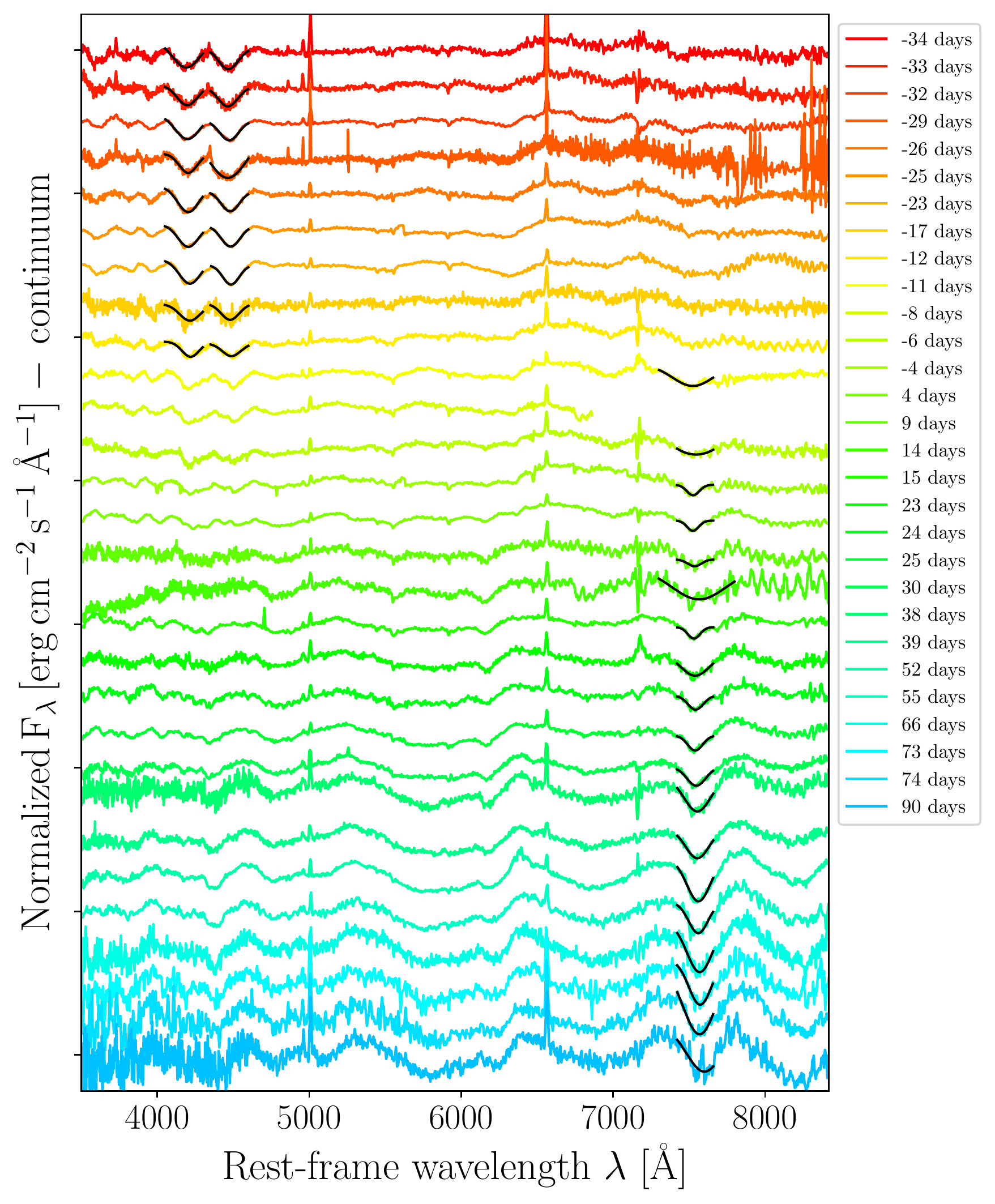}
    \caption{Normalized and continuum-subtracted spectra of SN~2018hti. A gaussian curve (black solid line) is fitted to the absorption minima of the \ion{O}{ii} $\lambda\lambda$ 4357,4650 and the \ion{O}{i} $\lambda\,7774$ features. {For the epochs in which more than one spectrum is available, we chose the spectrum with the {greatest} signal-to-noise ratio.} }
    \label{fig:vphot1}
\end{figure}

\begin{figure}
    \centering
    \includegraphics[width=0.5\textwidth]{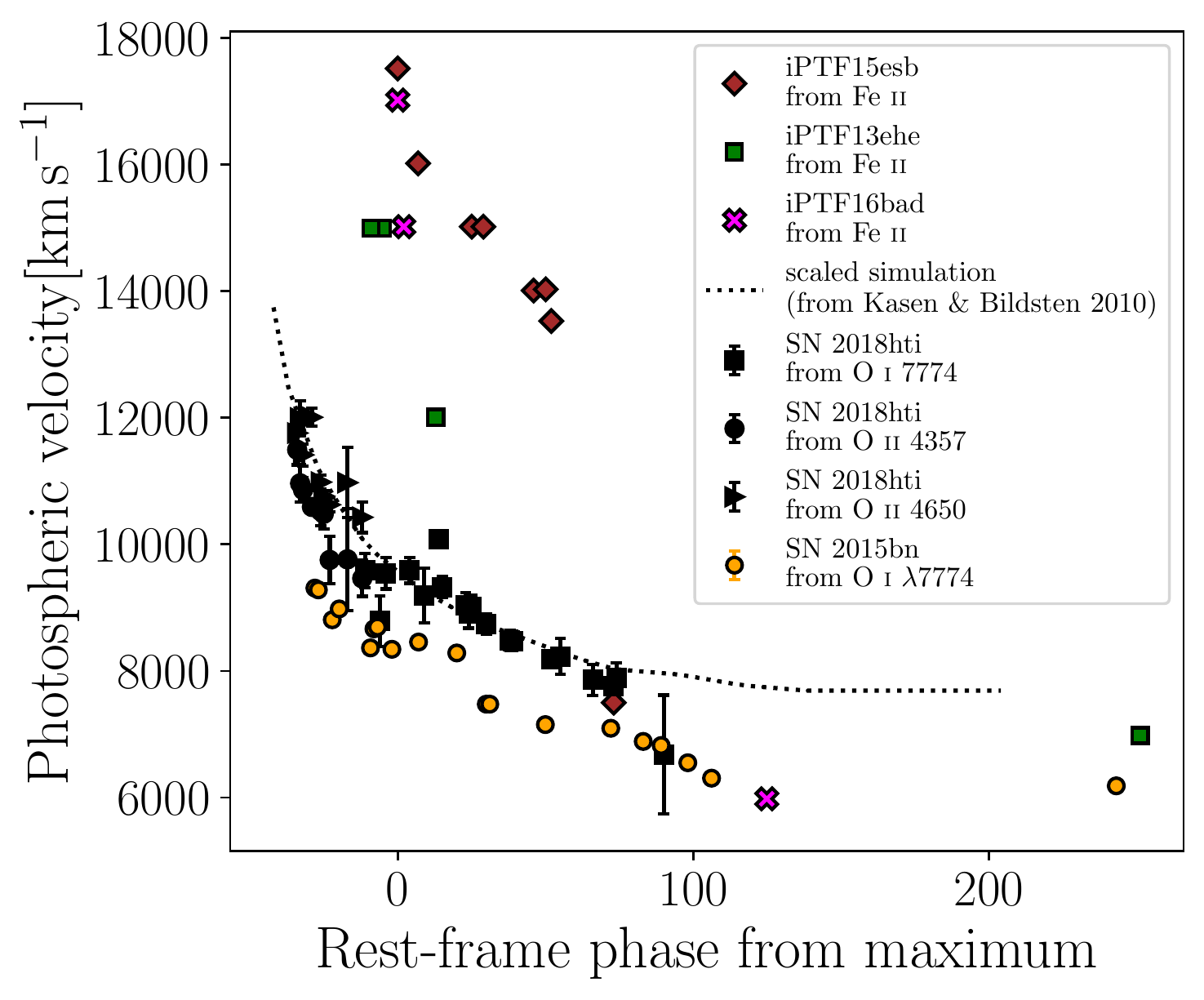}
    \caption{Photospheric velocities of SN~2018hti deduced from the absorption minima of O {\scriptsize II} $\lambda\lambda$ 4357,4650 (black dots and triangles) and O {\scriptsize I} $\lambda\,7774$ (black squares) P-Cygni profiles (see text). The evolution of the photospheric velocity of SN 2015bn \citep[deduced by the O {\scriptsize I}, data taken from][yellow dots]{nicholletal2016}, iPTF13ehe (green squares), iPTF15esb (brown diamonds), iPTF16bad (magenta crosses) \citep[deduced by the Fe {\scriptsize II},][]{yanetal2017b} are shown for comparison. {We also compared the photospheric velocity of SN~2018hti with the prediction of \citet{kasenandbildsten2010} for a magnetar-powered SN (black dotted line, see text).}}
    \label{fig:vphot}
\end{figure}

The photospheric velocity of SN~2018hti is measured via the O {\scriptsize II} $\lambda$ 4357, O {\scriptsize II} $\lambda$ 4650 and O {\scriptsize I} $\lambda$ 7774 P-Cygni absorption features present in the spectra. {In particular, it has been shown that the \ion{O}{i} $\lambda$ 7774 feature is a good tracer of the photosphere of the stripped envelope SNe \citep{dessartetal2015}. }The wavelengths corresponding to the absorption minima were inferred from a Gaussian fit of the absorption features (see Fig.~\ref{fig:vphot1}). {This method is marginally affected by the line blending \citep{jefferyandbranch1990} which can substantially bias the velocity measurements \citep{galyam2019b}. } The velocity evolution of SN~2018hti is shown in Fig.~\ref{fig:vphot} (right panel) in comparison with the photospheric velocities of SN~2015bn, iPTF13ehe, iPTF15esb and iPTF16bad (where the photospheric velocities of iPTF13ehe, iPTF15esb and iPTF16bad were retrieved based on the \ion{Fe}{ii} $\lambda$ 5169). In the case of SN~2018hti, after an initial very steep decline, the velocity evolution settles on an early plateau which starts $\sim22$ days before maximum luminosity and lasts $\sim$30 days. Overall, the photosphere of SN~2018hti recedes (in mass coordinates) similarly to SN~2015bn and both of them are much slower than the other SLSNe~I of the comparison sample. {Finally, we compared the photospheric-velocity evolution of SN~2018hti with the results of the numerical radiation hydrodynamic calculations of \citet[][see also their Fig.~2, bottom panel]{kasenandbildsten2010} for a magnetar-powered SN assuming a magnetic field $B_{\rm p}=0.5\times10^{14}$ G, an initial period $P_{\rm spin}=5$ ms, an ejecta mass $M_{\rm ejecta}=5\,\mathrm{M_\odot}$ and a kinetic energy $E_{\rm kin}=10^{51}$ erg. We scaled this solution by a factor 1.37 to almost perfectly fit the measured photospheric velocities of SN~2018hti (see also Sec.~\ref{sec:mosfit}).}

\subsection{Comparisons with other SLSNe~I}
\label{sec:comps}
\begin{figure*}
\centering
\includegraphics[width=0.8\textwidth]{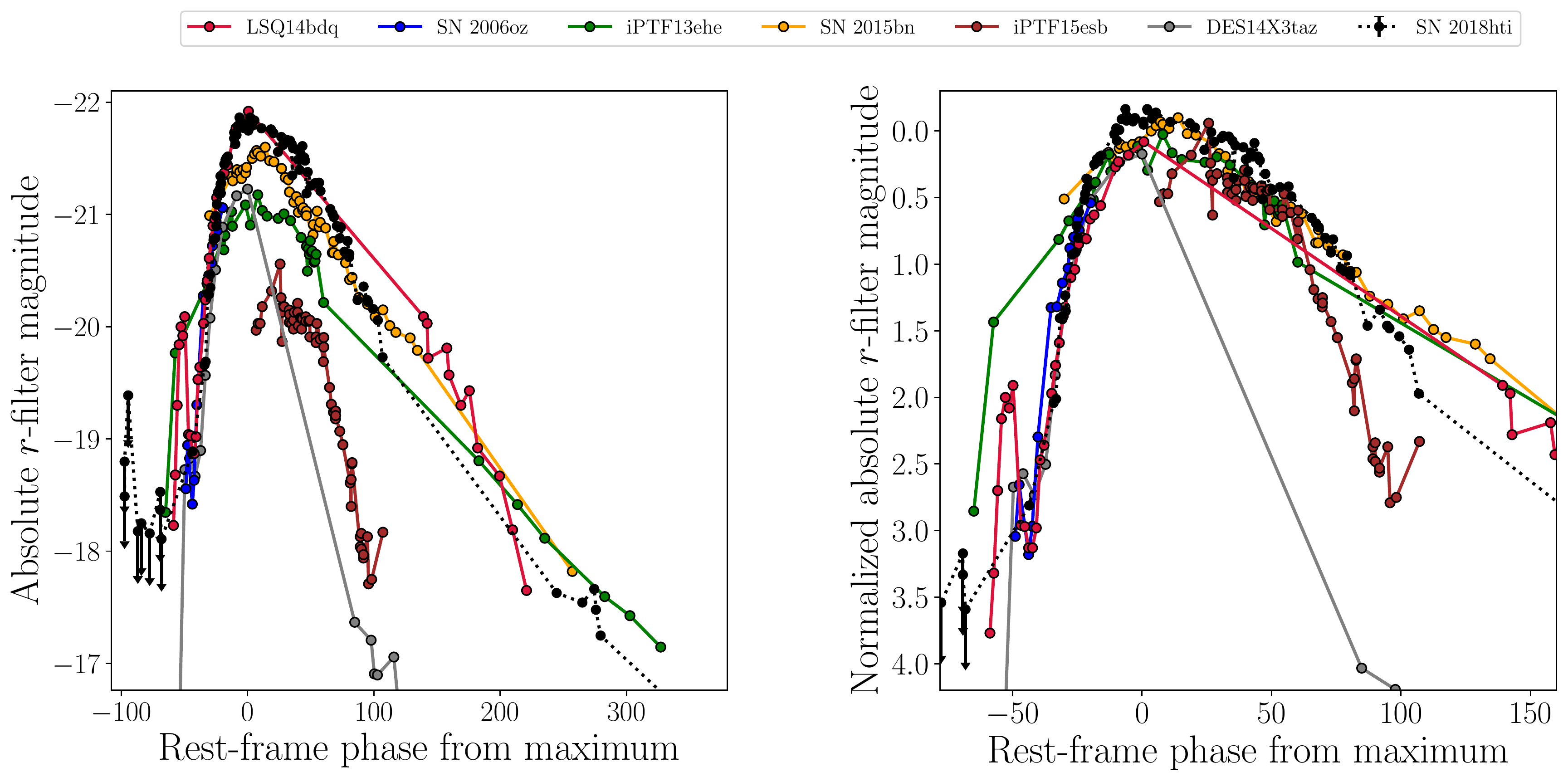}
\caption{Left panel: comparison of the $r$-filter absolute magnitude LC of SN~2018hti with those of LSQ14bdq (red dots), SN 2006oz (blue dots), iPTF13ehe (green dots), SN~2015bn (yellow dots), iPTF15esb (brown dots) {and DES14X3taz (gray dots)}. Right panel: same as the left panel, but with the LCs normalized to maximum luminosity. Absolute magnitudes were corrected for Galactic extinction as in Sec.~\ref{sec:photo} and calculated with the assumed cosmology in this work. Where $K$-correction values were not available, we assume a constant $K$-correction $2.5\log(1+z)$.}
\label{fig:lsq14bdq}
\end{figure*}

We compared the $r$-filter absolute magnitude LC of SN~2018hti with those of the comparison sample (see Fig.~\ref{fig:lsq14bdq}, left panel). These SLSNe~I show {an early} bump in their LCs {and/or} spectral signatures that likely involve some ejecta-CSM interaction {(i.e. a possible H$\alpha$ emergence)}. Interestingly, this sample shares similar evolutionary timescales (see Fig.~\ref{fig:lsq14bdq}, left panel) up to $\sim80$ days after maximum light, {even though} the absolute {peak} magnitude spans a range $>2$ mag.

\begin{figure*}
\centering\includegraphics[width=\textwidth]{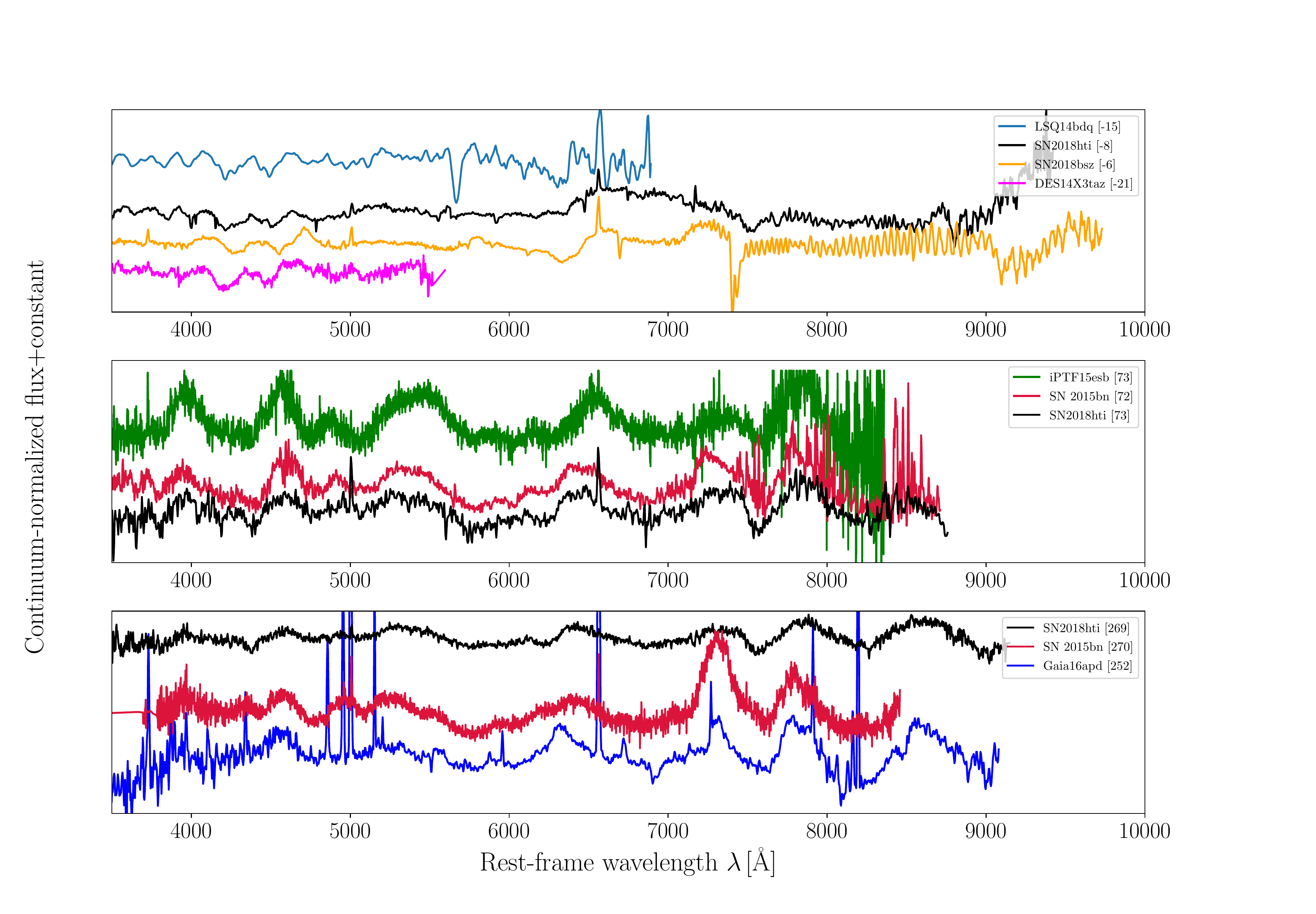}
\caption{Comparisons of three spectra of SN~2018hti at different phases with respect to maximum luminosity (the rest-frame phase from maximum luminosity is indicated among square brackets in the legend). Upper panel: comparison among SN~2018hti (8 days before maximum, black line), {SN~2018bsz} \citep[13 days before maximum, ][]{andersonetal2018}, LSQ14bdq \citep[15 days before maximum, light blue line, see][]{nicholletal2015} and DES14X3taz \citep[21 rest-frame days before maximum, magenta line, ][]{smithetal2016}. Middle panel: comparison among SN~2018hti (73 rest-frame days after maximum, black line), SN~2015bn (72 rest-frame days after maximum, red line) and iPTF15esb \citep[73 rest-frame days from maximum, green line, see][]{yanetal2017b}. Lower panel: comparison between SN~2018hti (269 rest-frame days after maximum, black line), SN~2015bn \citep[270 rest-frame days after maximum, red line, see][]{nicholletal2016} {and Gaia16apd \citep[252 rest-frame days after maximum, blue line, ][]{kangasetal2017}}. For a better visualization, the spectrum of LSQ14bdq was smoothed with a Savitzki-Golay filter {due to} its lower signal-to-noise ratio. The spectra of iPTF15esb, 2018bsz, LSQ14bdq, SN~2015bn {and of Gaia16apd} were obtained via \textsc{wiserep}.}
\label{fig:gelato}
\end{figure*}

Three representative spectra of SN~2018hti (at phases -8, +73, +269 days after maximum) {are} compared with the spectra of LSQ14bdq, {SN~2018bsz, DES14X3taz, Gaia16apd} and SN~2015bn (Fig.~\ref{fig:gelato}). The spectra of LSQ14bdq, {2018bsz, iPTF15esb, Gaia16apd, DES14X3taz} and SN~2015bn {are from} \textsc{wiserep}\footnote{\texttt{https://wiserep.weizmann.ac.il} .} \citep{yaronandgalyam2012}. The spectrum of SN~2018hti taken 8 days prior to maximum light is compared {with the spectra of LSQ14bdq (at a phase of -15 rest-frame days), SN~2018bsz (at a phase of -6 rest-frame days) and DES14X3taz (at a phase of -21 rest-frame days).} At these phases, the \ion{O}{ii} features in the blue {region} of the spectrum of SN~2018hti nicely match {those of LSQ14bdq, SN~2018bsz and DES14X3taz. However, in the earliest spectrum of SN~2018bsz the P-Cygni maximum of the \ion{O}{ii} $\lambda\,4650$ feature is likely affected by line blending with \ion{C}{ii} $\lambda\,4745$ \citep{andersonetal2018}}. At about 60-70 days after maximum luminosity, the spectrum of SN~2018hti is also similar to the spectrum of iPTF15esb, although the latter shows a more prominent \ion{Mg}{i}] $\lambda\,4571$ and broader Fe-group features at $\sim5500$ \AA{}. The remarkable resemblance between SN~2018hti and SN~2015bn at about $70$ days after maximum suggests that these SNe have similar ejecta velocities (see also Fig.~\ref{fig:vphot}) and chemical composition. Finally, the late/pseudo-nebular spectrum of SN~2018hti at 269 days after maximum is compared with the spectra of SN~2015bn {and Gaia16apd} at a phase +270 days {and +252 days}, respectively. The nebular emission features of SN~2015bn {and Gaia16apd} are {more strongly} developed compared to SN~2018hti, although they share some resemblance in the blue region of the spectrum. 
\subsection{Data interpretation}
\label{sec:dataint}
We considered two different scenarios to interpret the data of SN~2018hti: the magnetar and the ejecta-CSM interaction scenarios. To test the viability of the two hypotheses we modelled the multi-colour LCs of SN~2018hti with the Modular Open Source Fitter for Transients \citep[\textsc{mosfit}, ][]{guillochonetal2017soft,guillochonetal2018}. {We also }used the published radiative-transfer solutions of the \textsc{sumo} code \citep{jerkstrandetal2017} for the nebular emission of O-zone material in SLSNe I (see Sec.~\ref{sec:sumo}) as a guide for the interpretation of the pseudo-nebular spectrum of SN~2018hti. This allows us to constrain the mass of the progenitor of SN~2018hti.
\subsubsection{{The early boxy feature}}
\label{sec:boxy}
The flat-topped line profile of the emission feature at $\sim6500$ \AA{} could be suitably explained by emission inside an expanding shell of matter \citep{weiler2003,jerkstrand2017}. 
\begin{figure}
\centering
\includegraphics[width=0.45\textwidth]{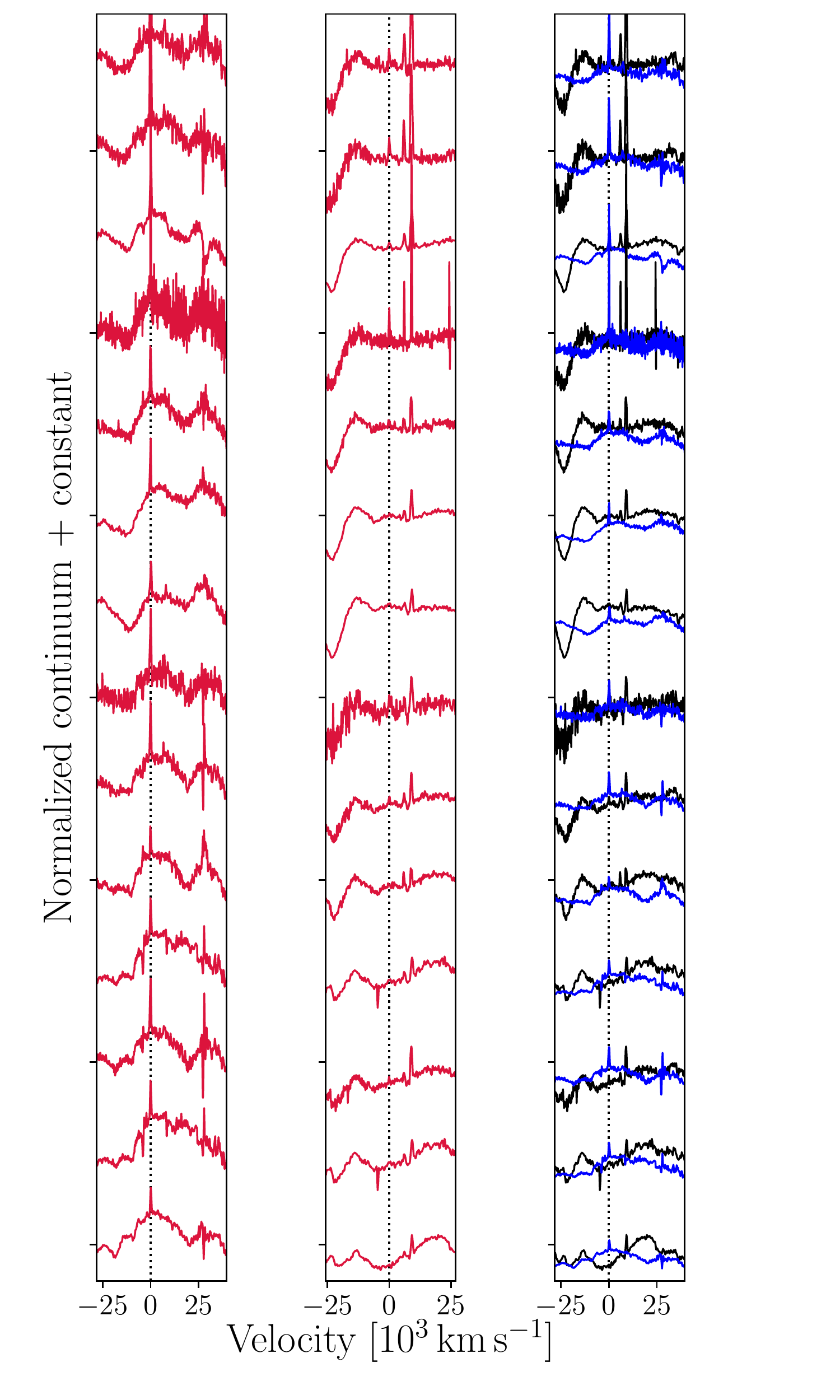}
\caption{Left panel: a close up of the H$\alpha$ region in the early spectral evolution of SN~2018hti (red solid lines). Central panel: same as in the left panel, but for the H$\beta$ region. Right panel: overlap of the H$\alpha$ (blue solid lines) and H$\beta$ (black solid lines) regions, where the H$\alpha$ region was superposed to that of H$\beta$. The dotted vertical black lines mark the rest-frame wavelength of H$\beta$ (middle panel) and of H$\alpha$ (left and right panel). The rest-frame phases of the spectra are labelled on the right side of the right panel.}
\label{fig:halpha}
\end{figure}
 The identification of this feature is not straightforward, and could be attributed either to H$\alpha$ or to \ion{C}{ii} $\lambda\,6580$. {To investigate this line identification, {we superimpose on top of the boxy feature the line profiles of} the possible H$\beta$ (where a tiny bump is present, see Fig.~\ref{fig:halpha}) and of \ion{C}{ii} $\lambda$ 9234 (see Fig.~\ref{fig:nirrrrcomp}). The comparison between the H$\alpha$ and H$\beta$ spectral regions is arduous because the H$\beta$ region is also {potentially contaminated} by other spectral features (such as \ion{O}{ii} and \ion{Fe}{ii}). On the other hand, the boxy line is well reproduced by the \ion{C}{ii} $\lambda$ 9234 feature in the IRTF+SpeX spectrum, suggesting that their flat profiles stem from the same matter shell and thus favouring a \ion{C}{ii} $\lambda$ 6580 identification.} 
 \begin{figure}
\includegraphics[width=0.55\textwidth]{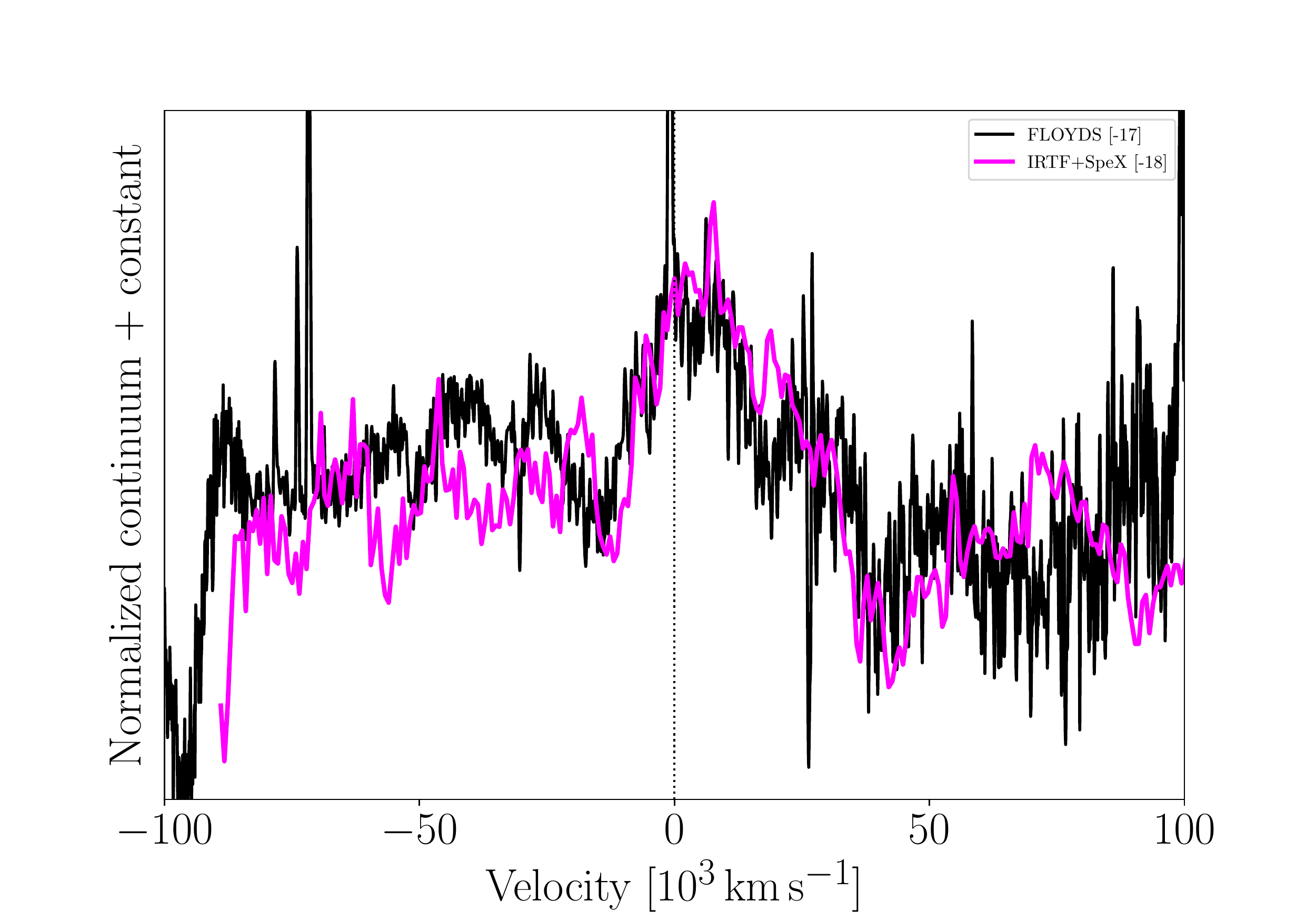}
\caption{Comparison between the continuum-normalized FLOYDS (black line) and IRTF+SpeX (magenta line) spectra of SN~2018hti at comparable phases. The optical and the NIR spectra are plotted in velocity coordinates with respect to $\lambda=6580$ \AA{} and $\lambda=9234$ \AA{}, respectively.}
\label{fig:nirrrrcomp}
\end{figure}
 Detailed radiative transfer calculations \citep[e.~g.][]{dessartetal2012,dessart2019} actually predict the presence of the \ion{C}{ii} $\lambda$ 6580 feature in the SLSNe~I spectra{, but} do not predict the boxy shape for the \ion{C}{ii} $\lambda$\,6580. {This suggests that the models may need to} more carefully account for dynamical effects such as the formation of a thick shell, which is expected from both from magnetar and CSM interaction scenario. 
\subsubsection{Light curves fits with \textsc{mosfit}}
\label{sec:mosfit}
 \textsc{mosfit} includes a number of models for different kinds of astronomical transients. In particular, those suitable for the SLSNe~I are the \textsc{csm} (CSM interaction powered), \textsc{csmni} (CSM interaction +$^{56}\mathrm{Ni}$-decay powered), the \textsc{slsn} and the \textsc{magnetar} (two implementations of the magnetar powered case, see later) and the \textsc{magni} (magnetar+$^{56}\mathrm{Ni}$-decay powered) models. We chose the \textsc{slsn} and the \textsc{csm} modules to fit the photometry of SN~2018hti, which respectively exploit the models introduced by \citet{inserraetal2013} and \citet{chatzopoulosetal2012}. Since \textsc{mosfit} takes as input the multiband LCs, it has to rebuild the pseudo-bolometric luminosities once a SED model has been assumed. We chose the \textsc{slsn} model since it accounts for the UV blanketing assuming an absorbed-blackbody model for the SED computation. {We excluded the $W1,W2$ magnitudes from the fit procedure since the MIR part of the SED could deviate from a single blackbody component at epochs which are not covered by our photometric dataset.}
 
\begin{figure}
    \centering
    \includegraphics[width=0.45\textwidth]{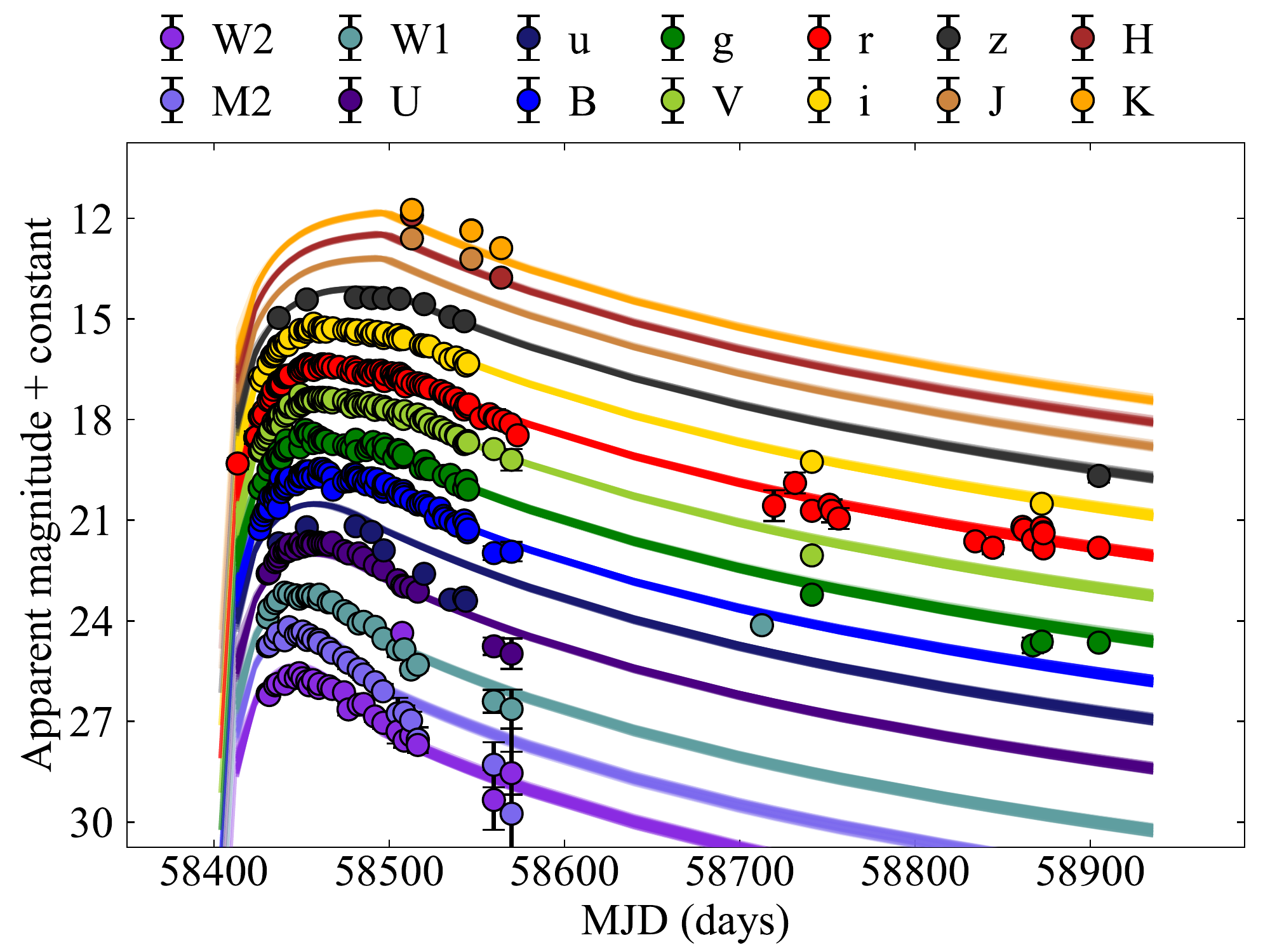}
    \caption{Best-fit \textsc{mosfit} synthetic LCs to the multiband photometry of SN~2018hti obtaned with the \textsc{slsn} model.}
    \label{fig:mosfit_mag}
\end{figure}

\begin{figure}
    \centering
    \includegraphics[width=0.45\textwidth]{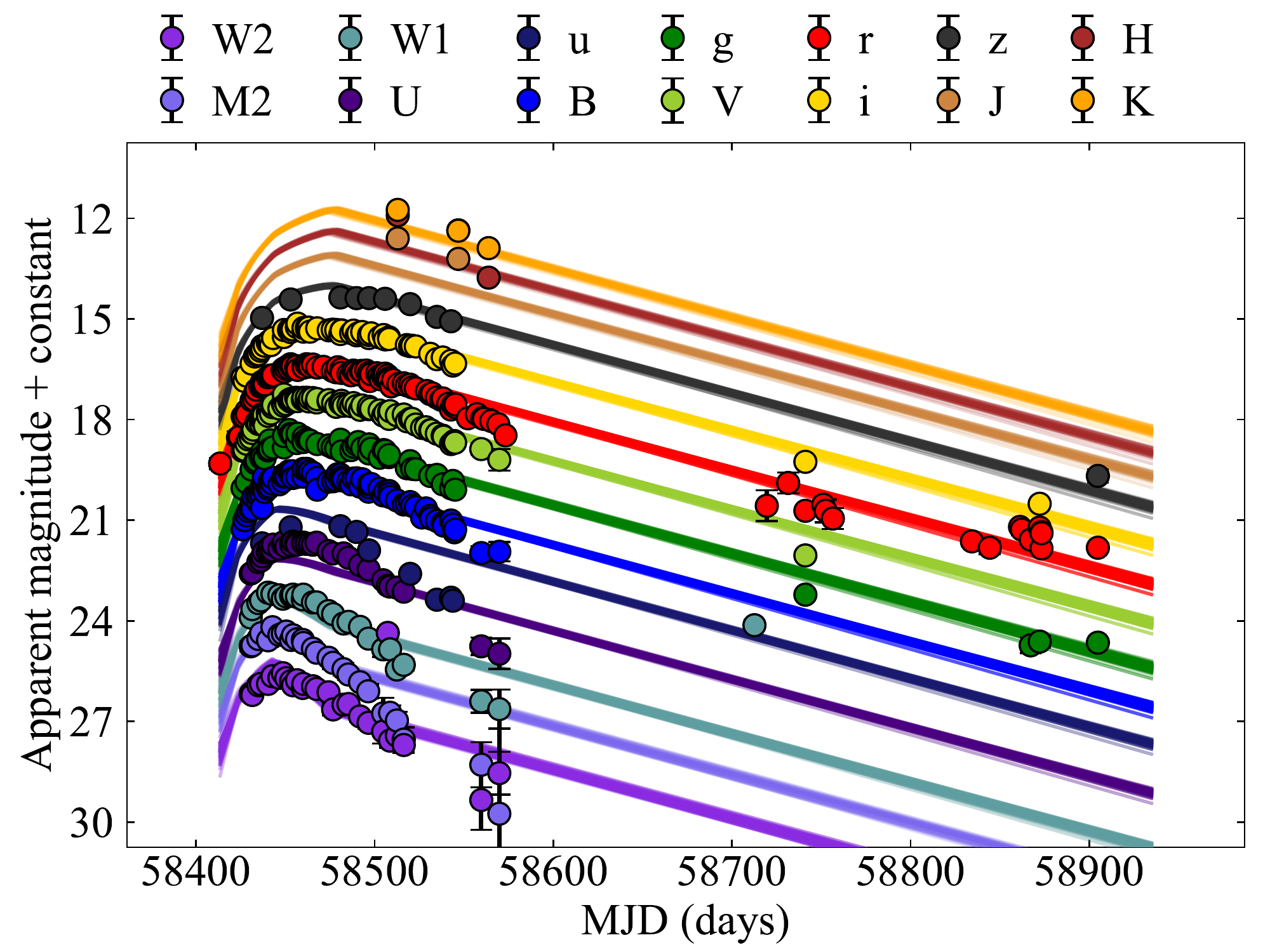}
    \caption{Same as in Fig.~\ref{fig:mosfit_mag}, but for the \textsc{csm} model.}
    \label{fig:mosfit_csm}
\end{figure}

\begin{figure*}
    \centering
    \includegraphics[width=0.9\textwidth]{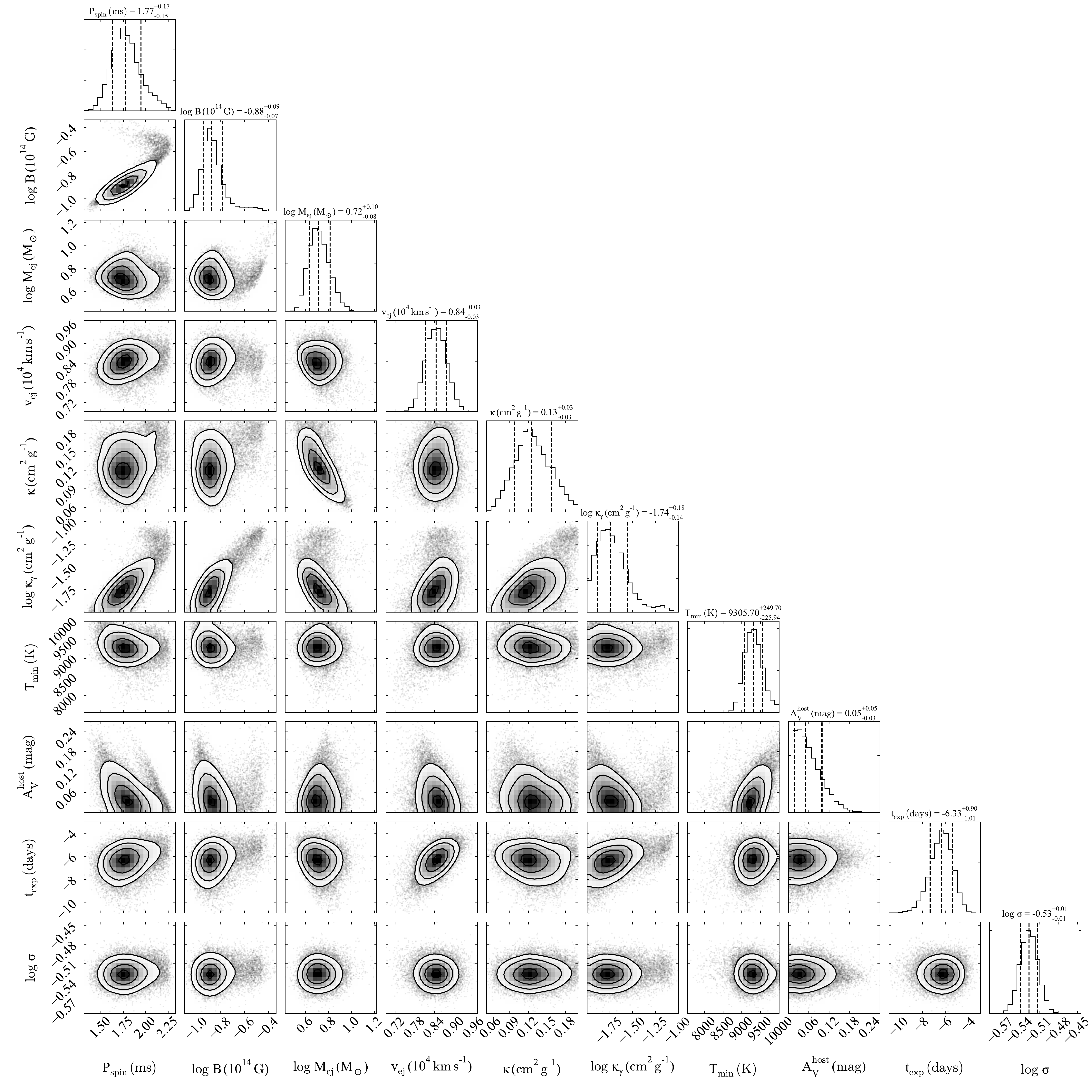}
    \caption{Corner plot with the best-fit parameters of the \textsc{mosfit} fit obtained with the \textsc{slsn} model.}
    \label{fig:corner_mag}
\end{figure*}

\begin{figure*}
    \centering
    \includegraphics[width=0.9\textwidth]{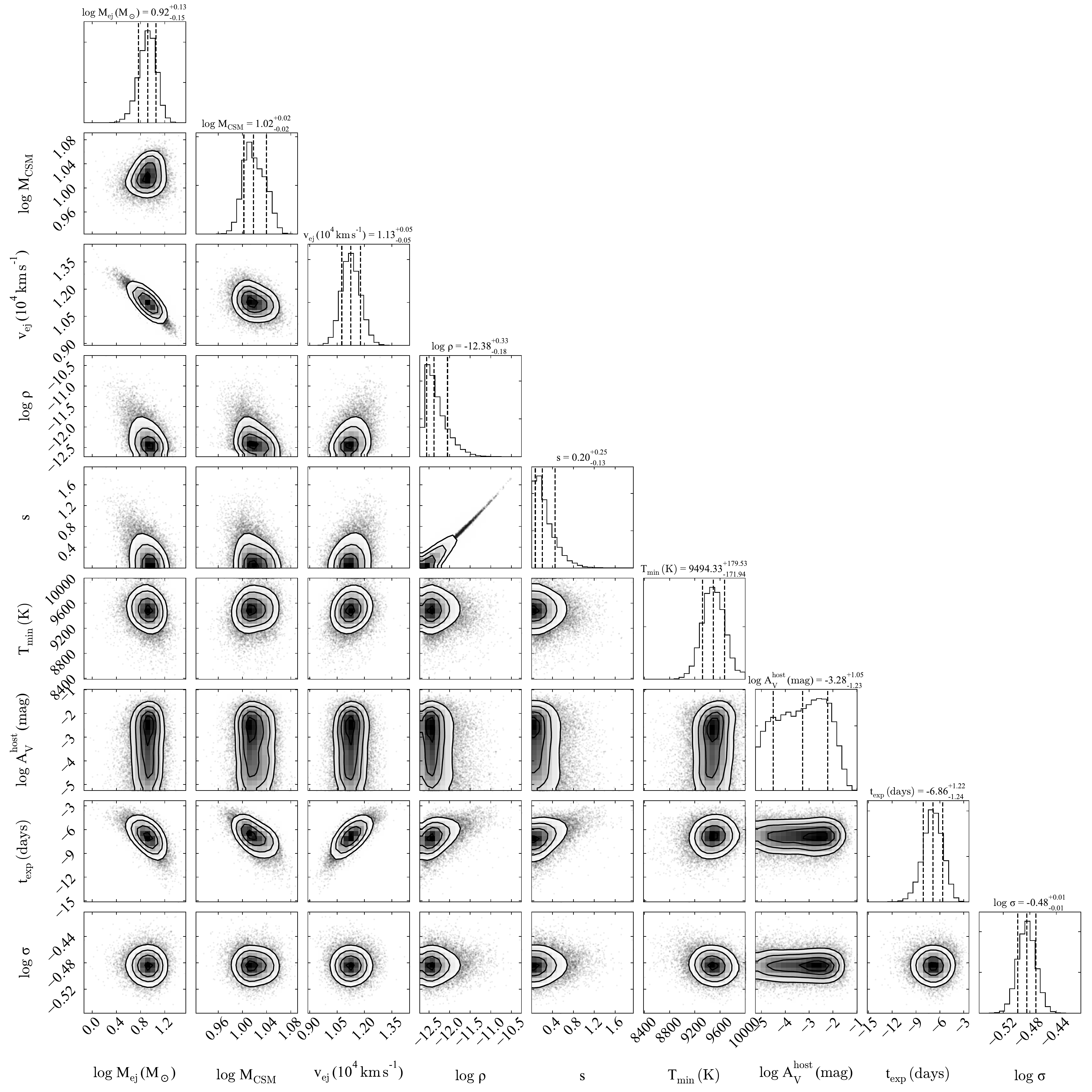}
    \caption{Same as Fig.~\ref{fig:corner_mag} but for the \textsc{csm} model.}
    \label{fig:corner_csm}
\end{figure*}

 Also, the \textsc{slsn} model includes constraints ensuring the energy conservation and that the ejecta do not become optically thin before 100 days after maximum, as not to contradict the late spectroscopic observations of the SLSNe \citep[see Sec. 3.8 in ][]{nicholletal2017}. The results of the fit procedures are shown in Fig.~\ref{fig:mosfit_mag},\ref{fig:mosfit_csm} and the corner plots showing the best-fit parameters are shown in Fig.~\ref{fig:corner_mag}, \ref{fig:corner_csm}. The \textsc{slsn} fit supports a magnetar engine with a polar magnetic field of $\sim1.3\times10^{13}\,\mathrm{G}$ and an initial period of $\sim1.8\,\mathrm{ms}$, for an ejecta mass $M_{\rm ejecta}\approx\,5.3\,\mathrm{M_\odot}$ (see Fig.~\ref{fig:corner_mag}), opacity $\kappa\approx0.1\,\mathrm{cm^2\,g^{-1}}$, gamma-ray opacity $\kappa_\gamma\approx0.02\,\mathrm{cm^2\,g^{-1}}$, {an average ejecta velocity $v_{\rm ej}\approx8500\,\mathrm{km\,s^{-1}}$} and a temperature floor $T_{\rm min}\approx9300\pm250\,\mathrm{K}$. {This corresponds to a kinetic energy {$E_{\rm kin}=3.7\times10^{51}\,\mathrm{erg}$.}} These results are absolutely reasonable for what is expected by the magnetar scenario for SLSNe~I \citep[e.~g.][]{nicholletal2017} and are in perfect agreement with the estimates of \citet{linetal2020a}.  {Except for the ejecta mass, the best-fit parameters for the magnetar case are quite different from those assumed in the calculations of \citet{kasenandbildsten2010}. This difference could possibly explain the need of the scaling factor 1.37 that we used in Sec. \ref{sec:vphot} to match the predicted photospheric velocity with the observed one. Moreover, the value of the kinetic energies required by both interpretations largely overcomes the maximum explosion energy that can be provided by a neutrino-driven mechanism during the core collapse \citep{sokerandgilkis2017,kaplanandsoker2020}. This energy budget might require the contribution of jets in the explosion of SN~2018hti. However, its negligible polarization degree \citep{lee2019} suggests that its explosion was nearly spherical, thus making this hypothesis less likely.} 

The \textsc{csm} fit of SN~2018hti instead requires the interaction of the SN ejecta with a mass of $\sim8.3\,\mathrm{M_\odot}$ and average velocity $v_{\rm ej}\approx1.1\times10^{4}\,\mathrm{km\,s^{-1}}$ with a CSM mass $M_{\rm CSM}\approx10.5\,\mathrm{M_\odot}$ and average density $\rho\approx4.1\times10^{-13}\,\mathrm{g\,cm^{-3}}$. {This corresponds to a kinetic energy $E_{\rm kin}=1.1\times10^{52}\,\mathrm{erg}$.} Also, for this model the predicted temperature floor reached by SN~2018hti is $T_{\rm min}\approx9500\pm180$ K. Both the predictions of $T_{\rm min}$ can be considered in agreement with what was deduced in Sec.~\ref{sec:tempevol}.
The best-fit slope of the CSM density profile $s\sim0.2$ seemingly favours a {shell-like CSM} with nearly-constant density \citep{chatzopoulosetal2013}. {The CSM interaction scenario may be disfavoured because of the of the absence of narrow/multicomponent features in the spectra (typical e.g. of SNe IIn) and because there was no signifcant detection in X-ray (see Sec.~\ref{sec:photo}). However, these arguments cannot rule out the CSM-interaction scenario for SN~2018hti if the CSM is highly asymmetric, e. g. if it has a disc-like geometry. In fact, if the CSM is not seen perfectly edge on, the optically thick ejecta may form a photosphere outside the CSM so that the ejecta CSM interaction takes place underneath it and the X-ray, UV photons can be reprocessed by further radiation-matter interactions \citep[as it was proposed by][for the peculiar SN II iPTF14hls]{andrewsandsmith2018}.}

\subsubsection{Interpretation of the nebular spectrum}
\label{sec:sumo}
The nebular spectrum of SN~2018hti taken 269 rest-frame days after maximum light was interpreted with \textsc{sumo} modelling \citep{jerkstrandetal2017}.
\begin{figure*}
    \centering
    \includegraphics[width=\textwidth]{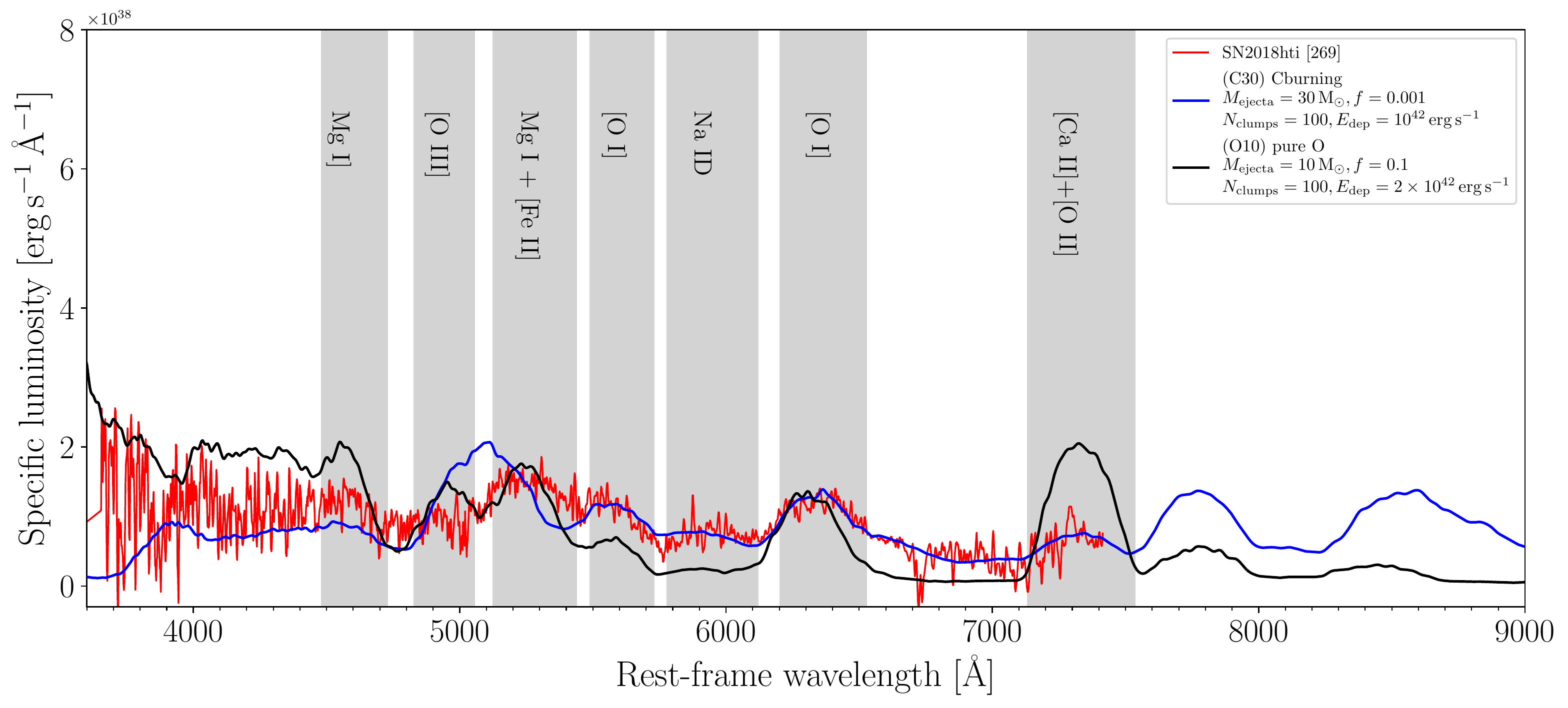}
    \caption{Comparison of the GTC+OSIRIS nebular spectrum (red line) with two outputs of the \textsc{sumo} numerical code (see text) for a full C-ashes model (blue line) and a pure-O composition (black line).}
    \label{fig:sumo}.
\end{figure*}
 The best-matching \textsc{sumo} models are built with a C-burning composition, $M_{\rm ejecta}=30\,\mathrm{M}_\odot$, {a filling factor} $f=0.001$, {an energy deposition} $E_{\rm dep}=10^{42}\,\mathrm{erg\,s^{-1}}$ and a pure-O abundance $M_{\rm ejecta}=10\,\mathrm{M}_\odot$, $f=0.1$, $E_{\rm dep}=2\times10^{42}\,\mathrm{erg\,s^{-1}}$. In the following, we will refer to them as C30 and O10 respectively. They are shown in Fig.~\ref{fig:sumo} with the pseudo-nebular spectrum of SN~2018hti. In particular, O10 better reproduces the bluer region of the spectrum (until $\sim5200$ \AA{}, see Fig.~\ref{fig:sumo}) whereas C30 better matches the redder region. Also, the best-matching spectra permit identification of other broad features in the spectrum, such as [\ion{O}{iii}] $\lambda\lambda\,4959,5007$, \ion{Mg}{i} $\lambda\,5180$+[\ion{Fe}{ii}] $\lambda\,5250$ and  [\ion{O}{i}] $\lambda\,5577$. We estimated the progenitor mass of SN~2018hti by measuring the flux emitted within the \ion{O}{i} $\lambda\,7774$ emission feature predicted by C30 using equations (7) and (8) of \citet{jerkstrandetal2017}. {The choice of C30 is motivated by the fact that it better describes the Oxygen features in the spectrum, as the [\ion{O}{i}] $\lambda\lambda\,6300,6364$ and [\ion{O}{i}] $\lambda\,5577$  features. The flux integrated within the \ion{O}{i} $\lambda\,7774$ feature gives $L_{7774}=2.25\times10^{40}\,\mathrm{erg\,s^{-1}}$. Hence we assumed $f=0.001$, the Oxygen mean molecular weight $\bar{A}=16$, an electron fraction $x_e=0.1$ \citep[][see their Sec. 4.2.1]{jerkstrandetal2017}, a maximum expansion velocity $V=8\,000\,\mathrm{km\,s^{-1}}$ {(we adopted for $V$ a value consistent with the velocity plateau at late times, see Fig.~\ref{fig:vphot})} and a recombination coefficient $\alpha^\mathrm{eff}(T)=2\times10^{-13}\,\mathrm{cm^3\,s^{-1}}$. Solving equation (7) of \citet{jerkstrandetal2017} for the electron density $n_e$}, this gives $n_e\sim1.28\times10^9\,\mathrm{cm^{-3}}$. Using this value in equation (8) in \citet{jerkstrandetal2017}, the O-zone mass is estimated to be $M_{\rm O-zone}\approx6.2\,\mathrm{M_\odot}$, which according to more recent models of stellar evolution of a single star corresponds to a progenitor mass $M_{\rm ZAMS}\approx40\,\mathrm{M_\odot}$ \citep{jerkstrandetal2017}. Similar consideration can be made for the O10 solution (corresponding to $f=0.1$ and $x_e\approx0.5$), which predicts a O-zone mass $M_{\rm O-zone}\approx10\,\mathrm{M_\odot}$ (for this solution we require $V\lesssim7\,000\,\mathrm{km\,s^{-1}}$ in order not to obtain $M_{\rm O-zone}>M_{\rm ejecta}$). In the latter case, the ejecta is expected to be much Mg-poorer compared to the C30 case.
 \begin{table}
     \centering
    \caption{Comparison of the ejecta masses of the best-matching \textsc{sumo} solutions with \textsc{mosfit} best-fit parameters.}
     \label{tab:comptab}
     \begin{tabular}{lll}
     \hline
     &ejecta mass $[\mathrm{M_\odot}]$\\
    \hline
          \textsc{sumo}&10-30&\\
          \textsc{mosfit csm}&8.32&\\
          \textsc{mosfit slsn}&5.25&\\
          \hline
     \end{tabular}
 \end{table}
 Another reason to favour the C30 model lies in its ejecta clump density. In fact, C30 is 300 times denser than O10\footnote{{The factor 300 comes from $(30\,\mathrm{M_\odot}/f_\mathrm{C30}) / (10\,\mathrm{M_\odot}/f_\mathrm{O10}) = 300$, where $f_\mathrm{C30}=0.001$ and $f_\mathrm{O10}=0.1$ are the clumping factors for the models C30 and O10, respectively.}}. This could be also the reason why no strong [\ion{O}{i}]$\lambda\,6300$ and [\ion{Ca}{ii}] + [\ion{O}{ii}] $\lambda\,7300$ emission is seen in the nebular spectrum, as it would emerge for higher-density models.
 
 {Finally, in Tab. \ref{tab:comptab} we summarized the ejecta-mass estimates obtained with the \textsc{mosfit} fits and the \textsc{sumo} nebular modelling. The \textsc{sumo} O10 solution apparently favours the CSM model since the ejecta mass used by O10 nearly reproduces that one estimated by the \textsc{mosfit} CSM fit, whereas the \textsc{mosfit} SLSN fit predicts an ejecta mass which is pretty lower than what is suggested by the \textsc{sumo} solutions. However, we warn the reader that the (single-zone) \textsc{sumo} solutions are computed for a phase of 400 days post-explosion, which is not the case for the pseudo-nebular spectrum of SN~2018hti. Hence, the density in the model is by a factor $(400/270)^{3}\simeq3.3$ lower than the corresponding case at 270 days. This biases a direct constrain on the ejecta density and mass. In addition, it is hard to believe that in the case of SN~2018hti the CSM interaction is acting as its major power source even if we interpret the modest \ion{C}{ii} boxy feature as a signature of the interaction with a CSM dense shell. According to the \textsc{mosfit} CSM fit, the predicted CSM mass is $\sim10.5\,\mathrm{M_\odot}$. We expect that the interaction with a similar amount of mass of CSM would cause strong spectral emissions as in the case of the type IIn SN~2008iy \citep{chugai2021} and SN~2010jl \citep{ofeketal2014}. However, as we mentioned earlier, a disk-like and dense CSM can hide the spectral signatures of CSM-interaction. Based on this considerations, we argue that the mechanism powering SN~2018hti could be either the spindown radiation from a millisecond magnetar with $B_{\rm p}\sim1.3\times10^{13}$ G and $P_{\rm spin}\sim1.8$ ms or the (buried) interaction of the ejecta with $\sim10\,\mathrm{M_\odot}$ of a disk-like CSM.}
\section{Conclusions}
In this work, we have presented the UV/optical/NIR photometry and the {NIR}/optical spectroscopy of the SLSN I SN 2018hti. It slowly rose for $\sim 50$ days towards a peak absolute {magnitude} of $\sim-21.7$ mag in the $r$-band. {Alongside this slow rise and extremely high luminosity}, the presence of the prominent \ion{O}{ii} absorptions in the pre-maximum/maximum spectra {identifies} this object as a (slow-evolving) SLSN~I. In the H$\alpha$ region, the early spectra show a flat-topped feature which we interpret as H$\alpha$. C-rich SLSNe~I spectra are predicted by magnetar- and a pair-instability driven radiative transfer calculations \citep{dessartetal2012,dessart2019}, but the boxy profile suggests that the feature could originate from the shock mediated interaction of the SN ejecta with a surrounding CSM. In addition, metallicity measurements via the host narrow emission lines are aligned with the low-metallicity paradigm of SLSNe~I. Finally, we estimated the physical parameters of the explosion, both in the magnetar and in the CSM-interaction scenarios, fitting synthetic LCs to the multicolor photometry of SN~2018hti with the \textsc{mosfit} tool. The model fits suggest that either interaction of a $8\,\mathrm{M_\odot}$ SN ejecta with $\sim10\,\mathrm{M_\odot}$ of CSM or the spindown radiation of a $B\sim1.3\times10^{13}$ G, $P_{\rm spin}\sim1.8$ ms magnetar could be the major power source for SN~2018hti.

We interpret the pseudo-nebular spectrum of SN~2018hti with synthetic spectra published by \citet{jerkstrandetal2017} for a SN Ic. We concluded that, assuming a single-star progenitor scenario for SN2018hti, the progenitor ZAMS mass was of $\sim40\,\mathrm{M}_{\odot}$. These findings help to unravel the origin of the complexities that often appear in SLSNe-I LCs \citep[e. g.][]{inserraetal2017}, finding a reasonable explanation in CSM-ejecta interaction. This sheds light on the nature of SLSNe~I progenitors.

The advent of the new generation, wide-field surveys such as the Legacy Survey of Space and Time at the Vera Rubin Observatory will contribute to broaden our knowledge about the SLSN astrophysics \citep{villaretal2018}.
\section*{Acknowledgements}
{We thank the anonymous referee for her/his valuable comments which improved the present work}. AF acknowledges Stephen Smarrt, Marica Branchesi, {Noam Soker} and Morgan Fraser for their suggestions and for interesting discussions. This work is based on observations made with the Nordic Optical Telescope, owned in collaboration by the University of Turku and Aarhus University, and operated jointly by Aarhus University, the University of Turku and the University of Oslo, representing Denmark, Finland and Norway, the University of Iceland and Stockholm University at the Observatorio del Roque de los Muchachos, La Palma, Spain, of the Instituto de Astrofisica de Canarias. The data presented here were obtained in part with ALFOSC, which is provided by the Instituto de Astrofisica de Andalucia (IAA) under a joint agreement with the University of Copenhagen and NOT. Based on observations collected at Copernico and Schmidt telescopes (Asiago, Italy) of the INAF - Osservatorio Astronomico di Padova. {MG is supported by the EU Horizon 2020 research and innovation programme under grant agreement No 101004719}. M.S. acknowledges the Infrared Telescope Facility, which is operated by the University of Hawaii under contract 80HQTR19D0030 with the National Aeronautics and Space Administration. N.E.R. acknowledges support from MIUR, PRIN 2017 (grant 20179ZF5KS). TMB was funded by the CONICYT PFCHA / DOCTORADOBECAS CHILE/2017-72180113. AJ acknowledges funding from the European Research Council (ERC) under the European Union's Horizon 2020 Research and Innovation Program (ERC Starting Grant. AR acknowledges support from ANID BECAS/DOCTORADO NACIONAL 21202412. Y.-Z. Cai is funded by China Postdoctoral Science Foundation (grant no. 2021M691821). {T.-W.C. acknowledges the EU Funding under Marie Sk\l{}odowska-Curie grant H2020-MSCA-IF-2018-842471. KM is funded by the EU H2020 ERC grant no. 758638. G. P.  is supported  by ANID - Millennium Science Initiative - ICN12\_009.}
We thank the staff at the different observatories for performing the observations. {Based on observations collected at the European organisation for astronomical research in the Southern Hemisphere, Chile, as part of ePESSTO+ (the advanced Public ESO Spectroscopic Survey for Transient Objects). ePESSTO+ observations were obtained under ESO program id 199.D-0143 (PI: Inserra)}.
Based on observations made with the Gran Telescopio Canarias (GTC), installed in the Spanish Observatorio del Roque de los Muchachos of the Instituto de Astrofísica de Canarias, in the island of La Palma. This work makes use of observations from the Las Cumbres Observatory network.  The LCO team is supported by NSF grants AST-1911225 and AST-1911151.
\section*{Data availability statement}
The data presented in this paper and listed in Appendix A are available in the online supplementary material. The spectra will be made public via \textsc{wiserep}. 



\bibliographystyle{mnras}
\bibliography{mnras_template_comments} 




\appendix

\section{Tables}
\begin{table*}
\centering
\caption{$uvw1,uvm2,uvw2$-filters observed (non $K$-corrected) {aperture} magnitudes (in AB system). Errors are in parentheses.}
\label{tab:18hti_uvottab}
\begin{tabular}{lp{30mm}llll}
\hline
MJD&r. f. phase&$uvw2$&$uvm2$&$uvw1$&instrument\\
&[days]&&&&\\
\hline
58430.65&-31.89&20.17(0.12)&19.76(0.14)&18.90(0.09)&\textit{Swift}/UVOT\\
58431.56&-31.03&20.21(0.12)&19.73(0.12)&18.65(0.08)&\textit{Swift}/UVOT\\
58434.92&-27.87&19.92(0.11)&19.54(0.12)&18.47(0.08&\textit{Swift}/UVOT\\
58436.44&-26.44&19.88(0.11)&19.37(0.11)&18.39(0.08)&\textit{Swift}/UVOT\\
58440.36&-22.74&19.88(0.17)&19.58(0.19)&18.16(0.11)&\textit{Swift}/UVOT\\
58442.82&-20.43&19.65(0.10)&19.19(0.10)&18.20(0.08)&\textit{Swift}/UVOT\\
58446.70&-16.77&19.70(0.10)&19.42(0.11)&18.24(0.07)&\textit{Swift}/UVOT\\
58448.62&-14.96&19.56(0.10)&19.38(0.11)&18.34(0.08)&\textit{Swift}/UVOT\\
58450.55&-13.14&19.71(0.10)&19.32(0.10)&18.27(0.07)&\textit{Swift}/UVOT\\
58453.25&-10.60&19.80(0.10)&19.57(0.13)&18.25(0.07)&\textit{Swift}/UVOT\\
58454.31&-9.60&19.92(0.11)&19.43(0.12)&18.22(0.07)&\textit{Swift}/UVOT\\
58456.57&-7.47&19.77(0.10)&19.55(0.12)&18.28(0.07)&\textit{Swift}/UVOT\\
58459.81&-4.42&20.00(0.11)&19.57(0.09)&18.36(0.07)&\textit{Swift}/UVOT\\
58460.14&-4.11&19.88(0.11)&19.66(0.10)&18.22(0.07)&\textit{Swift}/UVOT\\
58465.49&0.93&19.95(0.11)&19.83(0.11)&18.45(0.07)&\textit{Swift}/UVOT\\
58467.51&2.84&20.03(0.12)&19.95(0.11)&18.45(0.08)&\textit{Swift}/UVOT\\
58474.21&9.15&20.13(0.12)&20.10(0.11)&18.71(0.08)&\textit{Swift}/UVOT\\
58476.78&11.57&20.64(0.18)&20.25(0.18)&18.80(0.10)&\textit{Swift}/UVOT\\
58482.62&17.07&20.50(0.16)&20.44(0.18)&19.08(0.11)&\textit{Swift}/UVOT\\
58485.38&19.67&20.48(0.16)&20.60(0.16)&19.03(0.10)&\textit{Swift}/UVOT\\
58491.86&25.78&20.85(0.16)&20.83(0.14)&19.18(0.09)&\textit{Swift}/UVOT\\
58496.65&30.29&21.04(0.17)&21.10(0.21)&19.53(0.11)&\textit{Swift}/UVOT\\
58504.78&37.95&21.30(0.36)&21.76(0.47)&19.85(0.22)&\textit{Swift}/UVOT\\
58508.66&41.61&21.57(0.21)&21.74(0.22)&19.85(0.12)&\textit{Swift}/UVOT\\
58512.47&45.20&21.42(0.20)&21.97(0.26)&20.45(0.17)&\textit{Swift}/UVOT\\
58516.32&48.82&21.70(0.24)&22.54(0.37)&20.31(0.16)&\textit{Swift}/UVOT\\
58559.71&89.70&23.34(0.89)&23.29(0.67)&21.40(0.34)&\textit{Swift}/UVOT\\
58569.81&99.22&22.54(0.64)&24.75(3.28)&21.63(0.58)&\textit{Swift}/UVOT\\
\hline
\end{tabular}
\end{table*}
\begin{table*} 
\centering
\caption{$u,g,r,i,z$-filter observed (non $K$-corrected, {non S-corrected}) magnitudes (in AB system). Errors are in parentheses.}
\label{tab:18hti_ugriztab}
\begin{tabular}{lp{30mm}llllll}
\hline
MJD&r. f. phase&$u$&$g$&$r$&$i$&$z$&instrument\\
&[days]&&&&&\\
\hline
58413.54&-48.01&-&-&19.31(0.16)&-&-&ATLAS\\
58423.53&-38.60&-&-&18.54(0.19)&-&-&ATLAS\\
58424.54&-37.65&-&-&18.51(0.14)&-&-&ATLAS\\
58426.12&-36.16&-&18.05(0.01)&17.90(0.01)&17.76(0.02)&-&LCO+Sinistro\\
58427.17&-35.18&-&-&17.91(0.01)&17.86(0.01)&-&LCO+Sinistro\\
58428.33&-34.08&-&17.87(0.01)&17.73(0.01)&17.70(0.01)&-&LCO+Sinistro\\
58428.52&-33.90&-&-&17.85(0.03)&-&-&ATLAS\\
58431.14&-31.43&-&17.49(0.01)&17.43(0.01)&17.34(0.01)&-&LCO+Sinistro\\
58432.52&-30.13&-&-&17.41(0.04)&-&-&ATLAS\\
58433.03&-29.65&-&17.28(0.01)&17.21(0.01)&17.12(0.01)&-&LCO+Sinistro\\
58433.59&-29.12&-&-&17.30(0.12)&-&-&ATLAS\\
58433.94&-28.80&-&17.23(0.01)&17.11(0.01)&17.07(0.01)&-&LCO+Sinistro\\
58435.02&-27.77&-&-&-&17.05(0.02)&-&LCO+Sinistro\\
58436.15&-26.71&-&17.08(0.01)&17.02(0.01)&16.95(0.01)&-&LCO+Sinistro\\
58436.49&-26.39&-&-&16.97(0.04)&-&-&ATLAS\\
58436.85&-26.05&-&17.16(0.01)&16.94(0.01)&16.86(0.01)&-&LCO+Sinistro\\
58437.00&-25.91&18.19(0.03)&16.97(0.02)&16.86(0.01)&16.82(0.01)&16.97(0.01)&NOT+ALFOSC\\
58437.12&-25.80&-&17.10(0.01)&16.93(0.01)&16.89(0.01)&-&LCO+Sinistro\\
58437.51&-25.43&-&-&16.92(0.03)&-&-&ATLAS\\
58438.49&-24.51&-&17.15(0.01)&16.95(0.02)&16.80(0.01)&-&LCO+Sinistro\\
58440.24&-22.86&-&16.92(0.01)&16.75(0.01)&16.71(0.01)&-&LCO+Sinistro\\
58441.27&-21.89&-&16.61(0.03)&16.70(0.03)&16.66(0.02)&-&LCO+Sinistro\\
58441.30&-21.86&-&16.88(0.01)&16.71(0.01)&16.66(0.01)&-&LCO+Sinistro\\
58442.44&-20.78&-&-&16.69(0.05)&-&-&ATLAS\\
58443.20&-20.07&-&16.83(0.01)&16.68(0.01)&16.57(0.01)&-&LCO+Sinistro\\
58448.47&-15.10&-&-&16.52(0.03)&-&-&ATLAS\\
58449.23&-14.39&-&16.82(0.01)&-&16.55(0.01)&-&LCO+Sinistro\\
58449.25&-14.37&-&16.48(0.02)&16.48(0.01)&16.34(0.01)&-&LCO+Sinistro\\
58449.46&-14.17&-&-&16.57(0.01)&-&-&ATLAS\\
58450.47&-13.22&-&-&16.49(0.04)&-&-&ATLAS\\
58451.47&-12.28&-&-&16.41(0.02)&-&-&ATLAS\\
58452.10&-11.68&-&16.31(0.02)&16.47(0.02)&16.33(0.02)&-&LCO+Sinistro\\
58453.00&-10.84&17.71(0.03)&16.43(0.02)&16.33(0.01)&16.40(0.01)&16.42(0.01)&NOT+ALFOSC\\
58453.44&-10.42&-&-&16.42(0.03)&-&-&ATLAS\\
58453.58&-10.29&-&16.59(0.0)&16.39(0.01)&16.34(0.01)&-&LCO+Sinistro\\
58454.12&-9.78&-&16.43(0.01)&16.40(0.01)&16.34(0.01)&-&LCO+Sinistro\\
58456.09&-7.93&-&-&16.43(0.01)&-&-&LCO+Sinistro\\
58456.48&-7.56&-&16.69(0.01)&16.39(0.01)&16.33(0.01)&-&LCO+Sinistro\\
58456.86&-7.20&-&16.40(0.03)&16.40(0.02)&16.34(0.02)&-&Schmidt\\
58460.22&-4.04&-&16.51(0.01)&16.43(0.01)&16.39(0.01)&-&LCO+Sinistro\\
58460.23&-4.02&-&16.66(0.03)&16.39(0.02)&16.28(0.04)&-&LCO+Sinistro\\
58460.41&-3.85&-&-&16.45(0.02)&-&-&ATLAS\\
58461.93&-2.43&-&16.58(0.01)&16.33(0.01)&-&-&LCO+Sinistro\\
58462.09&-2.28&-&-&16.34(0.01)&16.28(0.01)&-&LCO+Sinistro\\
58462.09&-2.27&-&-&-&16.28(0.01)&-&LCO+Sinistro\\
58463.17&-1.26&-&16.61(0.01)&16.38(0.01)&16.26(0.01)&-&LCO+Sinistro\\
58463.84&-0.63&-&16.61(0.01)&16.41(0.02)&16.38(0.02)&-&LCO+Sinistro\\
58465.49&0.93&-&-&16.36(0.01)&-&-&ATLAS\\
58467.87&3.18&-&16.72(0.01)&16.43(0.01)&16.28(0.01)&-&LCO+Sinistro\\
58471.39&6.49&-&-&16.43(0.01)&-&-&ATLAS\\
58474.32&9.25&-&16.75(0.03)&-&16.34(0.04)&-&LCO+Sinistro\\
58476.55&11.35&-&16.82(0.01)&16.53(0.01)&16.32(0.01)&-&LCO+Sinistro\\
58478.52&13.21&-&16.85(0.01)&16.57(0.01)&16.33(0.01)&-&LCO+Sinistro\\
58479.38&14.02&-&-&16.44(0.02)&-&-&ATLAS\\
58480.86&15.41&17.68(0.01)&16.57(0.01)&-&16.37(0.01)&16.36(0.04)&1.82m+AFOSC\\
58481.45&15.96&-&16.67(0.0)&16.47(0.01)&16.34(0.01)&-&LCO+Sinistro\\
58481.56&16.07&-&16.94(0.02)&16.58(0.01)&16.35(0.01)&-&LCO+Sinistro\\
58485.36&19.65&-&-&16.64(0.03)&-&-&ATLAS\\
58486.86&21.07&-&16.89(0.02)&16.66(0.02)&16.42(0.02)&-&LCO+Sinistro\\
58488.17&22.30&-&16.79(0.01)&16.51(0.01)&16.35(0.01)&-&LCO+Sinistro\\
\hline
\end{tabular}
\end{table*}
\begin{table*}
\centering
\caption*{(continued).}
\label{tab:18hti_ugriztab_1}
\begin{tabular}{lp{30mm}llllll}
\hline
MJD&r. f. phase&$u$&$g$&$r$&$i$&$z$&instrument\\
&[days]&&&&&\\
\hline
58489.35&23.41&-&-&16.58(0.03)&-&-&ATLAS\\
58489.92&23.95&17.85(0.01)&16.71(0.01)&-)&16.41(0.01)&16.38(0.03)&1.82m+AFOSC\\
58490.82&24.79&-&16.66(0.01)&16.62(0.01)&16.43(0.01)&-&LCO+Sinistro\\
58492.36&26.25&-&-&16.55(0.03)&-&-&ATLAS\\
58493.07&26.91&-&16.83(0.01)&16.54(0.01)&16.39(0.01)&-&LCO+Sinistro\\
58495.35&29.07&-&-&16.54(0.01)&-&-&ATLAS\\
58496.81&30.44&-&16.99(0.02)&16.80(0.03)&16.57(0.04)&-&LCO+Sinistro\\
58496.87&30.50&18.40(0.01)&16.75(0.02)&16.57(0.01)&16.43(0.01)&16.38(0.01)&NOT+ALFOSC\\
58497.34&30.94&-&-&16.85(0.07)&-&-&ATLAS\\
58498.06&31.62&-&16.86(0.01)&16.61(0.01)&16.43(0.01)&-&LCO+Sinistro\\
58501.33&34.70&-&-&16.62(0.03)&-&-&ATLAS\\
58502.35&35.66&-&-&16.71(0.02)&-&-&ATLAS\\
58502.82&36.10&-&17.12(0.02)&-&16.55(0.02)&-&LCO+Sinistro\\
58503.37&36.62&-&-&16.80(0.05)&-&-&ATLAS\\
58504.38&37.57&-&-&16.66(0.06)&-&-&ATLAS\\
58505.06&38.21&-&16.90(0.02)&16.69(0.02)&16.48(0.02)&-&LCO+Sinistro\\
58505.25&38.39&-&-&16.69(0.10)&-&-&ATLAS\\
58505.87&38.98&18.63(0.01)&16.94(0.02)&16.59(0.02)&16.54(0.01)&16.40(0.01)&NOT+ALFOSC\\
58507.32&40.34&-&-&16.69(0.02)&-&-&ATLAS\\
58508.08&41.05&-&17.10(0.01)&16.72(0.01)&16.54(0.01)&-&LCO+Sinistro\\
58508.11&41.09&-&17.04(0.02)&16.78(0.05)&16.60(0.01)&-&LCO+Sinistro\\
58509.31&42.22&-&-&17.01(0.05)&-&-&ATLAS\\
58510.32&43.17&-&-&16.82(0.06)&-&-&ATLAS\\
58513.31&45.99&-&-&16.94(0.13)&-&-&ATLAS\\
58516.32&48.82&-&-&16.92(0.03)&-&-&ATLAS\\
58518.04&50.44&-&17.43(0.01)&16.93(0.01)&16.77(0.01)&-&LCO+Sinistro\\
58519.88&52.17&19.10(0.05)&17.26(0.01)&16.91(0.01)&16.74(0.01)&16.56(0.01)&NOT+ALFOSC\\
58521.04&53.26&-&17.46(0.01)&16.99(0.01)&16.80(0.01)&-&LCO+Sinistro\\
58523.06&55.17&-&17.47(0.01)&17.02(0.01)&16.82(0.01)&-&LCO+Sinistro\\
58526.00&57.94&-&-&17.18(0.01)&16.93(0.01)&-&Post Observatory\\
58529.29&61.04&-&-&17.15(0.04)&-&-&ATLAS\\
58531.15&62.79&-&17.66(0.02)&17.19(0.02)&17.04(0.02)&-&LCO+Sinistro\\
58532.24&63.82&-&-&17.22(0.08)&-&-&ATLAS\\
58534.86&66.29&19.87(0.02)&17.68(0.02)&17.30(0.01)&17.12(0.01)&16.94(0.02)&NOT+ALFOSC\\
58537.24&68.53&-&-&17.41(0.05)&-&-&ATLAS\\
58538.14&69.38&-&17.91(0.01)&17.31(0.01)&17.14(0.02)&-&LCO+Sinistro\\
58541.22&72.28&-&-&17.53(0.01)&-&-&ATLAS\\
58542.78&73.75&19.81(0.05)&18.02(0.03)&17.55(0.03)&17.14(0.04)&17.07(0.10)&1.82m+AFOSC\\
58543.80&74.72&19.90(0.17)&17.81(0.06)&17.43(0.05)&17.27(0.05)&-&Schmidt\\
58545.09&75.92&-&18.08(0.01)&17.58(0.01)&17.32(0.01)&-&LCO+Sinistro\\
58545.23&76.06&-&-&17.55(0.04)&-&-&ATLAS\\
58552.24&82.66&-&-&17.96(0.11)&-&-&ATLAS\\
58557.26&87.39&-&-&17.84(0.06)&-&-&ATLAS\\
58560.27&90.23&-&-&17.96(0.26)&-&-&ATLAS\\
58561.23&91.13&-&-&17.98(0.18)&-&-&ATLAS\\
58565.27&94.94&-&-&18.04(0.17)&-&-&ATLAS\\
58569.25&98.69&-&-&18.14(0.01)&-&-&ATLAS\\
58573.25&102.46&-&-&18.47(0.01)&-&-&ATLAS\\
58719.54&240.29&-&-&20.57(0.46)&-&-&ATLAS\\
58731.51&251.56&-&-&19.89(0.31)&-&-&1.2m+KeplerCam\\
58731.51&251.56&-&-&-&-&-&1.2m+KeplerCam\\
58741.15&260.64&-&21.21(0.07)&20.66(0.03)&20.49(0.02)&-&NOT+ALFOSC\\
58751.13&270.05&-&-&20.54(0.05)&-&-&OSIRIS\\
58752.55&271.39&-&-&20.72(0.34)&-&-&ATLAS\\
58756.52&275.13&-&-&20.95(0.31)&-&-&ATLAS\\
58834.37&348.47&-&-&21.63(0.14)&-&-&1.2m+KeplerCam\\
58838.38&352.25&-&-&$\gtrsim$20.93&-&-&1.2m+KeplerCam\\
58844.36&357.89&-&-&21.82(0.19)&-&-&ATLAS\\
58861.29&373.84&-&-&21.20(0.08)&$\gtrsim$21.08(0.01)&-&1.2m+KeplerCam\\
58862.27&374.76&-&-&21.28(0.17)&-&-&1.2m+KeplerCam\\
58867.25&379.45&-&22.73(0.23)&21.58(0.07)&-&-&1.2m+KeplerCam\\
\hline
\end{tabular}
\end{table*}
\begin{table*}
\caption*{(continued.)}
\begin{tabular}{lp{30mm}llllll}
\hline
MJD&r. f. phase&$u$&$g$&$r$&$i$&$z$&instrument\\
&[days]&&&&&\\
\hline
58872.28&384.19&-&22.62(0.18)&-&-&-&1.2m+KeplerCam\\
58872.29&384.20&-&-&21.22(0.07)&21.51(0.07)&-&1.2m+KeplerCam\\
58873.28&385.13&-&-&21.84(0.11)&-&-&1.2m+KeplerCam\\
58873.29&385.14&-&-&21.39(0.10)&-&-&1.2m+KeplerCam\\
58904.87&414.90&-&{$\gtrsim$}22.63(0.11)&{$\gtrsim$}21.74(0.07)&-&{$\gtrsim$}21.68(0.20)&NOT+ALFOSC\\
\hline
\end{tabular}
\end{table*}
\begin{table*} 
\centering
\caption{$U,B,V$-observed (non $K$-corrected, non S-corrected) magnitudes (in AB system).}
\label{tab:18hti_bvtab}
\begin{tabular}{lp{30mm}llll}
\hline
MJD&r. f. phase&$U$&$B$&$V$&instrument\\
&[days]&&&&\\
\hline
58426.12&-36.16&-&18.37(0.01)&17.90(0.02)&LCO+Sinistro\\
58427.15&-35.19&-&18.09(0.02)&17.87(0.01)&LCO+Sinistro\\
58428.31&-34.1&-&17.95(0.02)&17.69(0.01)&LCO+Sinistro\\
58429.29&-33.17&-&17.83(0.04)&17.56(0.01)&LCO+Sinistro\\
58430.66&-31.88&16.80(0.07)&17.76(0.08)&17.52(0.12)&\textit{Swift}/UVOT\\
58431.11&-31.46&-&17.56(0.01)&17.37(0.01)&LCO+Sinistro\\
58431.56&-31.03&16.77(0.07)&17.76(0.07)&17.38(0.10)&\textit{Swift}/UVOT\\
58432.11&-30.52&-&17.39(0.01)&17.27(0.01)&LCO+Sinistro\\
58433.92&-28.81&-&17.38(0.01)&17.06(0.01)&LCO+Sinistro\\
58434.92&-27.87&16.46(0.07)&17.57(0.07)&17.11(0.1)&\textit{Swift}/UVOT\\
58435.0&-27.79&-&17.36(0.01)&-&LCO+Sinistro\\
58436.14&-26.72&-&17.18(0.02)&16.96(0.01)&LCO+Sinistro\\
58436.44&-26.44&16.41(0.06)&17.31(0.06)&17.12(0.09)&\textit{Swift}/UVOT\\
58436.85&-26.05&-&17.73(0.03)&16.97(0.07)&LCO+Sinistro\\
58436.99&-25.92&-&17.16(0.02)&16.82(0.01)&NOT+ALFOSC\\
58437.1&-25.81&16.29(0.04)&17.1(0.02)&16.87(0.01)&LCO+Sinistro\\
58437.92&-25.04&-&17.17(0.02)&-&LCO+Sinistro\\
58438.01&-24.96&16.45(0.05)&-&16.87(0.01)&LCO+Sinistro\\
58438.01&-24.96&-&17.08(0.01)&-&LCO+Sinistro\\
58438.03&-24.94&16.19(0.01)&16.82(0.02)&16.83(0.04)&LCO+Sinistro\\
58438.49&-24.51&-&17.25(0.03)&17.14(0.09)&LCO+Sinistro\\
58440.21&-22.88&16.16(0.07)&-&16.77(0.03)&LCO+Sinistro\\
58440.36&-22.74&16.14(0.09)&17.15(0.09)&16.87(0.14)&\textit{Swift}/UVOT\\
58441.27&-21.89&-&17.09(0.04)&16.64(0.02)&LCO+Sinistro\\
58441.28&-21.88&16.02(0.03)&16.93(0.02)&16.66(0.01)&LCO+Sinistro\\
58442.82&-20.43&16.11(0.06)&17.0(0.06)&16.61(0.08)&\textit{Swift}/UVOT\\
58443.18&-20.09&16.14(0.03)&16.84(0.02)&16.53(0.02)&LCO+Sinistro\\
58446.7&-16.77&16.06(0.06)&17.02(0.06)&16.65(0.08)&\textit{Swift}/UVOT\\
58448.62&-14.96&15.92(0.06)&16.93(0.06)&16.50(0.07)&\textit{Swift}/UVOT\\
58449.22&-14.4&15.92(0.05)&16.75(0.04)&16.23(0.02)&LCO+Sinistro\\
58449.23&-14.39&-&16.8(0.03)&16.57(0.02)&LCO+Sinistro\\
58450.56&-13.13&16.00(0.06)&16.78(0.05)&16.52(0.07)&\textit{Swift}/UVOT\\
58452.1&-11.68&-&16.64(0.08)&16.40(0.01)&LCO+Sinistro\\
58452.99&-10.84&-&16.66(0.01)&16.36(0.01)&NOT+ALFOSC\\
58453.25&-10.6&15.92(0.06)&16.75(0.05)&16.45(0.07)&\textit{Swift}/UVOT\\
58453.56&-10.31&15.96(0.04)&16.71(0.01)&16.41(0.01)&LCO+Sinistro\\
58454.12&-9.78&-&16.63(0.03)&16.37(0.01)&LCO+Sinistro\\
58454.31&-9.6&15.85(0.06)&16.83(0.05)&16.38(0.07)&\textit{Swift}/UVOT\\
58456.07&-7.94&15.88(0.01)&-&16.34(0.01)&LCO+Sinistro\\
58456.57&-7.47&15.94(0.06)&16.80(0.05)&16.38(0.07)&\textit{Swift}/UVOT\\
58456.85&-7.21&-&16.5(0.02)&16.26(0.02)&AFOSC\\
58456.85&-7.21&-&-&16.41(0.01)&Schmidt\\
58459.81&-4.42&15.88(0.06)&16.77(0.06)&16.38(0.07)&\textit{Swift}/UVOT\\
58460.14&-4.11&15.88(0.06)&16.79(0.06)&16.38(0.08)&\textit{Swift}/UVOT\\
58460.19&-4.06&15.91(0.04)&16.57(0.01)&16.33(0.01)&LCO+Sinistro\\
58461.92&-2.43&15.91(0.03)&16.58(0.03)&16.36(0.02)&LCO+Sinistro\\
58462.07&-2.29&15.91(0.02)&16.56(0.01)&16.33(0.01)&LCO+Sinistro\\
58463.17&-1.25&-&16.64(0.01)&16.45(0.01)&LCO+Sinistro\\
58463.82&-0.64&15.92(0.03)&16.63(0.02)&16.36(0.01)&LCO+Sinistro\\
58465.49&0.93&15.94(0.06)&16.68(0.06)&16.34(0.07)&\textit{Swift}/UVOT\\
58467.51&2.84&15.88(0.06)&16.83(0.06)&16.47(0.08)&\textit{Swift}/UVOT\\
58467.88&3.18&-&17.18(0.07)&16.38(0.02)&LCO+Sinistro\\
58474.21&9.15&16.13(0.06)&16.94(0.06)&16.44(0.07)&\textit{Swift}/UVOT\\
58474.3&9.23&-&-&16.41(0.07)&LCO+Sinistro\\
58476.78&11.57&16.20(0.07)&16.93(0.07)&16.49(0.10)&\textit{Swift}/UVOT\\
58478.52&13.21&-&16.91(0.03)&16.61(0.03)&LCO+Sinistro\\
58479.3&13.94&-&16.72(0.03)&-&LCO+Sinistro\\
58480.85&15.4&-&16.75(0.02)&16.48(0.02)&AFOSC\\
58481.43&15.95&-&16.82(0.02)&16.43(0.01)&LCO+Sinistro\\
58481.56&16.07&-&16.90(0.06)&16.52(0.01)&LCO+Sinistro\\
\hline
\end{tabular}
\end{table*}
\begin{table*}
\centering
\caption*{(continued.)}
\begin{tabular}{lp{30mm}llll}
\hline
MJD&r. f. phase&$U$&$B$&$V$&instrument\\
&[days]&&&&\\
\hline
58482.62&17.07&16.20(0.07)&16.87(0.07)&16.57(0.09)&\textit{Swift}/UVOT\\
58485.38&19.67&16.33(0.07)&16.92(0.07)&16.54(0.09)&\textit{Swift}/UVOT\\
58486.86&21.07&-&17.02(0.05)&16.59(0.03)&LCO+Sinistro\\
58488.15&22.28&-&16.85(0.02)&16.51(0.01)&LCO+Sinistro\\
58489.92&23.95&-&16.86(0.03)&16.55(0.02)&AFOSC\\
58491.86&25.78&16.55(0.07)&17.09(0.06)&16.69(0.08)&\textit{Swift}/UVOT\\
58493.06&26.91&-&16.87(0.02)&16.57(0.01)&LCO+Sinistro\\
58496.65&30.29&16.68(0.07)&17.24(0.07)&16.67(0.09)&\textit{Swift}/UVOT\\
58496.81&30.44&-&17.11(0.02)&16.66(0.02)&LCO+Sinistro\\
58496.86&30.49&-&17.03(0.01)&16.60(0.01)&NOT+ALFOSC\\
58498.05&31.61&-&17.02(0.04)&16.67(0.02)&LCO+Sinistro\\
58502.81&36.09&-&17.15(0.03)&16.72(0.01)&LCO+Sinistro\\
58504.78&37.95&17.0(0.14)&17.35(0.12)&16.86(0.16)&\textit{Swift}/UVOT\\
58505.87&38.98&-&17.25(0.01)&16.78(0.01)&NOT+ALFOSC\\
58508.07&41.05&-&17.23(0.01)&16.77(0.01)&LCO+Sinistro\\
58508.11&41.09&-&17.32(0.03)&16.97(0.07)&LCO+Sinistro\\
58508.66&41.61&17.19(0.08)&17.40(0.07)&16.73(0.08)&\textit{Swift}/UVOT\\
58512.47&45.2&17.22(0.08)&17.50(0.07)&16.85(0.08)&\textit{Swift}/UVOT\\
58516.32&48.82&17.34(0.09)&17.62(0.07)&16.83(0.08)&\textit{Swift}/UVOT\\
58518.04&50.44&-&17.59(0.03)&17.04(0.02)&LCO+Sinistro\\
58519.86&52.16&-&17.57(0.01)&16.97(0.01)&NOT+ALFOSC\\
58521.03&53.26&-&17.66(0.02)&16.97(0.03)&LCO+Sinistro\\
58523.06&55.17&-&17.71(0.02)&17.14(0.01)&LCO+Sinistro\\
58526.57&58.48&-&18.01(0.06)&17.23(0.04)&Post Observatory\\
58528.76&60.54&-&17.80(0.13)&17.19(0.05)&Schmidt\\
58531.14&62.79&-&17.97(0.04)&17.32(0.03)&LCO+Sinistro\\
58531.61&63.23&-&18.04(0.04)&17.30(0.02)&Post Observatory\\
58534.85&66.28&-&18.14(0.02)&17.41(0.01)&NOT+ALFOSC\\
58538.12&69.36&-&18.19(0.02)&17.48(0.01)&LCO+Sinistro\\
58538.64&69.85&-&18.25(0.03)&17.49(0.02)&Post Observatory\\
58542.78&73.75&-&18.18(0.04)&17.68(0.03)&AFOSC\\
58543.8&74.71&-&18.31(0.08)&17.64(0.03)&Schmidt\\
58544.65&75.51&-&18.45(0.03)&17.62(0.02)&Post Observatory\\
58545.07&75.91&-&18.38(0.04)&17.69(0.05)&LCO+Sinistro\\
58559.72&89.71&18.97(0.26)&19.07(0.21)&17.86(0.17)&\textit{Swift}/UVOT\\
58569.81&99.22&19.19(0.45)&19.03(0.29)&18.18(0.32)&\textit{Swift}/UVOT\\
58741.14&260.64&-&-&21.05(0.03)&NOT+ALFOSC\\
\hline
\end{tabular}
\end{table*}
\begin{table*} 
\centering
\caption{$J,H,K_{\rm s}$-observed (non $K$-corrected) magnitudes (in AB system). Errors are in parentheses.}
\label{tab:18hti_jhktab}
\begin{tabular}{lp{30mm}llll}
\hline
MJD&r. f. phase&$J$&$H$&$K_\mathrm{s}$&instrument\\
&[days]&&&&\\
\hline
58512.96&45.56&15.60(0.01)&15.910.01)&16.75(0.01)&NOT+NOTCam\\
58546.87&77.51&16.21(0.02)&16.360.02)&17.37(0.03)&NOT+NOTCam\\
58563.85&93.50&-&17.77(0.04)&17.89(0.04)&NOT+NOTCam\\
\hline
\end{tabular}
\end{table*}
\begin{table*}
\caption{$W1,W2$-observed (non $K$-corrected) magnitudes (in AB system). Errors are in parentheses.}
\label{tab:18hti_wise}
\begin{tabular}{lp{30mm}lll}
\hline
MJD&r. f. phase&$W1$&$W2$&instrument\\
&[days]&&&\\
\hline
58507.39&40.04&17.95(0.07)&18.33(0.13)&WISE\\
58712.64&233.69&19.13(0.16)&$\gtrsim18.61$&WISE\\
\hline
\end{tabular}
\end{table*}
\begin{table*}
\caption{S-corrections for Schmidt and AFOSC filters (Asiago observatory).}
\label{tab:scorrasiago}
\begin{tabular}{ccccccc}
\hline
MJD&$B$&$g$&$V$&$r$&$i$\\
\hline
58430.25&-0.02&0.01&0.018&0.113&0.053\\
58437.02&-0.019&0.014&0.022&0.122&0.048\\
58437.14&-0.017&0.014&0.019&0.127&0.06\\
58440.23&-0.027&0.012&0.018&0.101&0.064\\
58452.97&-0.014&0.018&0.022&0.122&0.05\\
58456.79&-0.004&0.019&0.021&0.115&-0.206\\
58460.21&-0.011&0.021&0.022&0.133&0.052\\
58468.16&-0.009&0.026&0.028&0.139&0.073\\
58480.81&-0.015&0.027&0.024&0.139&0.077\\
58489.87&-0.019&0.029&0.024&0.139&0.076\\
58491.12&-0.022&0.027&0.018&0.146&0.094\\
58496.89&-0.033&0.03&0.022&0.147&0.093\\
58505.86&-0.037&0.03&0.022&0.152&0.109\\
58519.89&-0.058&0.031&0.023&0.152&0.105\\
58522.04&-0.042&0.022&0.019&0.147&0.105\\
58534.87&-0.067&0.033&0.027&0.148&0.124\\
58542.81&-0.08&0.032&0.022&0.146&0.117\\
58543.80&-0.086&0.028&0.02&0.14&0.117\\
58751.14&-0.052&0.044&-0.013&0.124&-0.154\\
\hline
\end{tabular}
\end{table*}
\begin{table*}
\caption{S-corrections for Sinistro (LCO).}
\label{tab:scorrlco}
\begin{tabular}{ccccccc}
\hline
MJD&$B$&$g$&$V$&$r$&$i$\\
\hline
58430.25&-0.009&0.003&0.009&0.004&-0.005\\
58437.02&-0.008&0.004&0.003&0.007&-0.002\\
58437.14&-0.008&0.005&0.011&0.005&-0.001\\
58440.23&-0.009&0.004&0.009&-0.007&-0.0\\
58452.97&-0.005&0.008&-0.002&0.008&-0.002\\
58456.79&-0.005&0.012&-0.012&0.014&-0.101\\
58460.21&-0.004&0.011&-0.001&0.014&-0.007\\
58468.16&-0.002&0.015&-0.012&0.018&-0.004\\
58480.81&-0.001&0.015&-0.011&0.025&0.002\\
58489.87&-0.002&0.018&-0.015&0.029&-0.001\\
58491.12&-0.001&0.015&-0.004&0.034&-0.002\\
58496.89&-0.002&0.017&-0.014&0.044&0.002\\
58505.86&-0.003&0.016&-0.012&0.049&0.005\\
58519.89&-0.006&0.016&-0.018&0.058&0.005\\
58522.04&-0.006&0.01&-0.007&0.048&0.002\\
58534.87&-0.009&0.016&-0.024&0.054&0.005\\
58542.81&-0.01&0.016&-0.022&0.058&0.009\\
58543.80&-0.009&0.011&-0.018&0.052&0.01\\
58751.14&-0.015&0.033&-0.04&0.062&-0.013\\
\hline
\end{tabular}
\end{table*}
\begin{table*}
\caption{S-corrections for NOT filters.}
\label{tab:scorrnot}
\begin{tabular}{ccccccc}
\hline
MJD&$B$&$g$&$V$&$r$&$i$\\
\hline
58430.25&-0.013&0.001&0.008&0.011&-0.008\\
58437.02&-0.013&-0.004&0.005&0.013&-0.011\\
58437.14&-0.014&-0.002&0.011&0.015&-0.013\\
58440.23&-0.015&-0.005&0.007&0.002&0.011\\
58452.97&-0.01&-0.001&0.002&0.014&-0.01\\
58456.79&-0.008&0.007&-0.005&0.019&-0.313\\
58460.21&-0.008&0.001&0.004&0.022&-0.019\\
58468.16&-0.006&0.001&-0.004&0.022&0.011\\
58480.81&-0.008&-0.001&-0.003&0.026&0.031\\
58489.87&-0.012&-0.001&-0.005&0.028&0.033\\
58491.12&-0.009&-0.004&0.003&0.036&0.034\\
58496.89&-0.013&-0.009&-0.003&0.041&0.052\\
58505.86&-0.015&-0.011&-0.002&0.045&0.07\\
58519.89&-0.023&-0.018&-0.005&0.047&0.068\\
58522.04&-0.019&-0.012&0.001&0.041&0.065\\
58534.87&-0.027&-0.023&-0.007&0.043&0.089\\
58542.81&-0.031&-0.026&-0.007&0.05&0.09\\
58543.80&-0.034&-0.028&-0.004&0.047&0.086\\
58751.14&-0.021&0.003&-0.025&0.068&-0.251\\
\hline
\end{tabular}
\end{table*}
\begin{table*}
\caption{S-corrections for \textit{Swift}/UVOT.}
\label{tab:scorrswift}
\begin{tabular}{ccccccc}
\hline
MJD&$B$&$V$\\
\hline
58430.25&-0.017&0.009\\
58437.02&-0.018&0.008\\
58437.14&-0.017&0.019\\
58440.23&-0.02&0.006\\
58452.97&-0.014&0.011\\
58456.79&-0.014&0.005\\
58460.21&-0.01&0.019\\
58468.16&-0.008&0.012\\
58480.81&-0.007&0.016\\
58489.87&-0.005&0.02\\
58491.12&-0.005&0.034\\
58496.89&-0.009&0.029\\
58505.86&-0.011&0.034\\
58519.89&-0.018&0.033\\
58522.04&-0.014&0.034\\
58534.87&-0.025&0.033\\
58542.81&-0.026&0.031\\
58543.80&-0.025&0.035\\
58751.14&-0.024&0.049\\
\hline
\end{tabular}
\end{table*}
\begin{table*}
    \centering
        \caption{Estimated uncertainties $\Delta S_{\rm corr}$ for the filters $u,U,z,J,H,K_{\rm s}$ (for each instrument) divided in two temperature ranges (see text). }
        \label{tab:scorrerr}
    \begin{tabular}{lll}
    
    \hline
         &{$5000\,\mathrm{K}<T<10000\,\mathrm{K}$}&{$10000\,\mathrm{K}<T<20000\,\mathrm{K}$}  \\
         \hline
         {NOT+ALFOSC/NOTCam}&\begin{tabular}{l}{$\Delta S_{{\rm corr,}u}=0.30$}\\{$\Delta S_{{\rm corr,}z}=0.03$}\\{$\Delta S_{{\rm corr,}J}=0.1$}\\{$\Delta S_{{\rm corr,}H}=0.1$}\\{$\Delta S_{{\rm corr,}K_{\rm s}}=0.1$}\end{tabular}&\begin{tabular}{l}{$\Delta S_{{\rm corr,}u}=0.20$}\\{$\Delta S_{{\rm corr,}z}=0.01$}\\{$\Delta S_{{\rm corr,}J}=0.1$}\\{$\Delta S_{{\rm corr,}H}=0.1$}\\{$\Delta S_{{\rm corr,}K_{\rm s}}=0.1$}\end{tabular}\\
         \hline
         {Schmidt/AFOSC}&\begin{tabular}{l}{$\Delta S_{{\rm corr,}u}=0.3$}\\{$\Delta S_{{\rm corr,}z}=0.3$}\end{tabular}&\begin{tabular}{l}{$\Delta S_{{\rm corr,}u}=0.2$}\\{$\Delta S_{{\rm corr,}z}=-0.001$}\end{tabular}\\
         \hline
         {LCO+Sinistro}&\begin{tabular}{l}{$\Delta S_{{\rm corr,}z}=0.001\,\mathrm{mag}$}\end{tabular}&\begin{tabular}{l}{$\Delta S_{{\rm corr,}z}=0.005\,\mathrm{mag}$}\end{tabular}\\
         \hline
         {\textit{Swift}/UVOT}&\begin{tabular}{l}{$\Delta S_{{\rm corr,}U}=0.2\,\mathrm{mag}$}\end{tabular}&\begin{tabular}{l}{$\Delta S_{{\rm corr,}U}=0.05\,\mathrm{mag}$}\end{tabular}\\
         \hline
    \end{tabular}
    \label{tab:deltascorr}
\end{table*}
\begin{table*}
\caption{$K$-corrections expressed in magnitudes.} 
\label{tab:18hti_kcorr}
\begin{tabular}{lllllllllllllll}
\hline
rest-frame phase &$uvw2$&$uvm2$&$uvw1$&$u$&$U$&$B$&$g$&$V$&$r$&$i$&$z$&$J$&$H$&$K_\mathrm{s}$\\
$[\mathrm{days}]$&filter&filter&filter&filter&filter&filter&filter&filter&filter&filter&filter&filter&filter&filter\\
\hline
-32.36&-0.106&-0.098&0.031&0.191&0.191&0.009&-0.008&0.030&-0.002&0.043&-0.206&-0.154&-0.030&0.402\\
-25.98&-0.100&-0.103&0.037&0.166&0.173&-0.015&-0.022&0.028&0.004&0.068&-0.203&-0.159&-0.030&0.148\\
-25.87&-0.102&-0.105&0.037&0.165&0.173&-0.011&-0.016&0.038&0.005&0.131&-0.203&-0.159&-0.030&0.146\\
-22.96&-0.083&-0.055&0.080&0.129&0.137&-0.020&-0.023&0.028&0.013&-0.012&-0.202&-0.147&-0.033&0.089\\
-10.96&-0.064&-0.052&0.122&0.150&0.155&-0.013&-0.025&0.014&0.003&0.071&-0.214&-0.129&-0.032&-0.022\\
-7.36&-0.085&-0.060&0.146&0.127&0.135&0.020&-0.028&-&-&-&-0.191&-0.142&-0.038&-0.011\\
-4.15&-0.040&0.022&0.148&0.136&0.152&-0.004&-0.020&0.021&0.008&0.097&-0.195&-0.129&-0.039&-0.010\\
3.35&-0.030&0.038&0.194&0.129&0.131&-0.017&-0.037&-0.019&-0.052&0.034&-0.200&-0.127&-0.038&-0.017\\
15.27&-0.046&-0.002&0.276&0.080&0.084&-0.03&-0.043&-0.005&-0.023&0.006&-0.171&-0.139&-0.046&0.003\\
23.81&-0.042&0.080&0.281&0.083&0.100&-0.045&-0.058&-0.032&-0.049&-0.010&-0.164&-0.132&-0.051&0.018\\
24.99&-0.048&0.081&0.272&0.106&0.129&-0.05&-0.062&-0.001&-0.007&0.000&-0.168&-0.133&-0.050&0.015\\
30.42&-0.057&0.082&0.251&0.181&0.215&-0.076&-0.088&-0.032&-0.066&-0.058&-0.16&-0.122&-0.054&0.031\\
38.87&-0.067&0.097&0.288&0.178&0.218&-0.086&-0.096&-0.035&-0.054&-0.091&-0.138&-0.116&-0.067&0.076\\
52.09&-0.079&0.140&0.285&0.209&0.255&-0.126&-0.127&-0.051&-0.056&-0.068&-0.144&-0.100&-0.075&0.167\\
54.13&-0.094&0.121&0.274&0.228&0.277&-0.111&-0.114&-0.052&-0.063&-0.089&-0.147&-0.100&-0.073&0.179\\
66.21&-0.182&0.003&0.254&0.215&0.270&-0.159&-0.155&-0.069&-0.041&-0.103&-0.136&-0.116&-0.064&0.251\\
73.69&-0.229&-0.071&0.227&0.215&0.268&-0.171&-0.154&-0.059&-0.069&-0.114&-0.110&-0.096&-0.077&0.344\\
74.63&-0.235&-0.080&0.219&0.221&0.272&-0.192&-0.165&-0.048&-0.035&-0.107&-0.100&-0.098&-0.082&0.369\\
\hline
\end{tabular}
\end{table*}
\begin{table*} 
\centering
\caption{Logarithm of the bolometric luminosities integrated over the $uvw2,uvm2,uvw1,U,B,g,V,r,i,z,J,H,K_\mathrm{s},W1,W2$.}
\label{tab:18hti_blc}
\begin{tabular}{p{30mm}l}
\hline
rest-frame phase&$\log_{10}L_\mathrm{bol}$\\
$[\mathrm{days}]$&\\
\hline
-48.01&43.30(0.04)\\
-38.60&43.61(0.04)\\
-37.65&43.62(0.04)\\
-36.16&43.86(0.04)\\
-35.17&43.87(0.04)\\
-34.08&43.94(0.04)\\
-33.90&43.91(0.04)\\
-31.43&44.07(0.04)\\
-30.13&44.10(0.04)\\
-29.65&44.14(0.04)\\
-29.12&44.14(0.04)\\
-28.80&44.17(0.04)\\
-26.71&44.22(0.04)\\
-26.39&44.22(0.04)\\
-26.05&44.22(0.04)\\
-25.91&44.26(0.04)\\
-25.80&44.24(0.04)\\
-25.43&44.25(0.04)\\
-24.50&44.24(0.04)\\
-22.85&44.29(0.04)\\
-21.88&44.32(0.04)\\
-21.87&44.32(0.04)\\
-21.20&44.31(0.04)\\
-20.78&44.32(0.04)\\
-20.08&44.34(0.04)\\
-15.10&44.35(0.04)\\
-14.37&44.38(0.04)\\
-14.17&44.37(0.04)\\
-13.22&44.38(0.04)\\
-12.27&44.39(0.04)\\
-11.68&44.39(0.04)\\
-10.83&44.40(0.04)\\
-10.42&44.38(0.04)\\
-10.30&44.39(0.04)\\
-9.78&44.40(0.04)\\
-7.93&44.38(0.04)\\
-7.56&44.38(0.04)\\
-7.20&44.38(0.04)\\
-4.04&44.39(0.04)\\
-4.02&44.39(0.04)\\
-3.85&44.39(0.04)\\
-2.42&44.40(0.04)\\
-2.28&44.39(0.04)\\
-1.25&44.38(0.04)\\
-0.63&44.38(0.04)\\
0.93&44.38(0.04)\\
3.18&44.35(0.04)\\
6.49&44.34(0.04)\\
11.35&44.29(0.04)\\
13.21&44.28(0.04)\\
14.02&44.31(0.04)\\
15.96&44.30(0.04)\\
16.08&44.27(0.04)\\
19.66&44.26(0.04)\\
21.07&44.24(0.04)\\
22.30&44.27(0.04)\\
23.41&44.26(0.04)\\
24.80&44.24(0.04)\\
26.25&44.24(0.04)\\
26.92&44.24(0.04)\\
29.07&44.21(0.04)\\
30.44&44.16(0.04)\\
\hline
\end{tabular}
\end{table*}
\begin{table*}
\centering
\caption*{(continued).}
\begin{tabular}{p{30mm}l}
\hline
rest-frame phase&$\log_{10}L_\mathrm{bol}$\\
$[\mathrm{days}]$&\\
\hline
30.50&44.21(0.04)\\
30.94&44.17(0.04)\\
31.62&44.19(0.04)\\
34.70&44.17(0.04)\\
35.66&44.15(0.04)\\
36.62&44.13(0.04)\\
37.58&44.15(0.04)\\
38.22&44.14(0.04)\\
38.39&44.15(0.04)\\
38.98&44.15(0.04)\\
40.35&44.14(0.04)\\
41.05&44.12(0.04)\\
41.09&44.10(0.04)\\
42.22&44.07(0.04)\\
43.17&44.09(0.04)\\
45.99&44.05(0.04)\\
48.82&44.03(0.04)\\
50.44&44.01(0.04)\\
52.18&44.02(0.04)\\
53.27&44.00(0.04)\\
55.17&43.98(0.04)\\
61.04&43.92(0.04)\\
62.79&43.90(0.04)\\
63.82&43.89(0.04)\\
66.29&43.86(0.04)\\
68.53&43.84(0.04)\\
69.38&43.84(0.04)\\
72.28&43.81(0.04)\\
73.75&43.79(0.04)\\
74.71&43.79(0.04)\\
75.92&43.77(0.04)\\
76.06&43.77(0.04)\\
82.67&43.67(0.04)\\
87.40&43.65(0.04)\\
90.23&43.61(0.04)\\
91.14&43.60(0.04)\\
94.94&43.58(0.04)\\
98.69&43.55(0.04)\\
102.46&43.47(0.04)\\
240.29&42.62(0.04)\\
251.57&42.77(0.04)\\
260.64&42.54(0.04)\\
270.05&42.57(0.04)\\
271.39&42.52(0.04)\\
275.13&42.45(0.04)\\
348.47&42.15(0.04)\\
357.89&42.08(0.04)\\
373.84&42.24(0.04)\\
374.76&42.22(0.04)\\
379.45&42.12(0.04)\\
384.20&42.23(0.04)\\
385.13&42.18(0.04)\\
385.13&42.02(0.04)\\
385.14&42.17(0.04)\\
414.90&$\gtrsim$42.01(0.04)\\
\hline
\end{tabular}
\end{table*}
\begin{table*}
\centering
\begin{threeparttable}
\caption{Spectra in Fig.~\ref{fig:spec}.} 
\label{tab:18hti_sfo}
\begin{tabular}{lllll}
\hline
MJD&rest-frame phase&instrumental set-up [grism/grating]&resolution\\
&[days]&&[\AA{}]\\
\hline
58428.57&-34&LCO+FLOYDS&15.5\\
58429.57&-33&LCO+FLOYDS&15\\
58430.25&-32&NTT+EFOSC2 [gr13]&18\\
58433.19&-29&HET+LRS2&-\\
58436.35&-26&LCO+FLOYDS($^{*}$)&15\\
58437.02&-26&NOT+ALFOSC [gr4]&14\\
58437.14&-25&NTT+EFOSC2 [gr11+gr16] &15\\
58440.23&-23&NTT+EFOSC2 [gr11+gr16]&18\\
58445.41&-18&IRTF+SpeX&-\\
58446.52&-17&LCO+FLOYDS &-\\
58450.28&-13&2.4m Hiltner+OSMOS&-\\
58450.33&-13&HET+LRS2&-\\
58451.45&-12&LCO+FLOYDS&15\\
58452.97&-11&NOT+ALFOSC [gr4]&14\\
58456.79&-8&1.82m+AFOSC [VPH6+VPH7]&15\\
58458.49&-6&LCO+FLOYDS&15\\
58460.21&-4&NTT+EFOSC2 [gr11+gr16]&14\\
58468.16&4&NTT+EFOSC2 [gr11+gr16]&14\\
58468.47&4&LCO+FLOYDS&15\\
58474.46&9&LCO+FLOYDS&15\\
58479.55&14&LCO+FLOYDS&19\\
58480.81&15&1.82m+AFOSC [VPH6+VPH7]&20\\
58488.32&23&LCO+FLOYDS&15\\
58489.87&24&1.82m+AFOSC [VPH6+VPH7]&17\\
58491.12&25&NTT+EFOSC2 [gr11+gr16]&14\\
58495.32&29&LCO+FLOYDS&15\\
58496.89&30&NOT+ALFOSC [gr4]&15\\
58504.37&38&LCO+FLOYDS&14\\
58505.86&39&NOT+ALFOSC [gr4]&18\\
58519.89&52&NOT+ALFOSC [gr4]&14\\
58522.05&55&NTT+EFOSC2 [gr]&15\\
58534.87&66&NOT+ALFOSC [gr4]&14\\
58542.81&73&1.82m+AFOSC [VPH6+VPH7]&14\\
58543.80&74&1.82m+AFOSC [VPH6+VPH7]&15\\
58546.28&77&LCO+FLOYDS&15\\
58559.23&90&LCO+FLOYDS&15\\
58751.14&269&GTC+OSIRIS [R1000B]&9\\
\hline
\end{tabular}
   \begin{tablenotes}
      \small
      \item ($^{*}$) This spectrum was not included in Fig.~\ref{fig:spec} because of its poor signal-to-noise ratio, but it will be made available within the online dataset (see the Data Availability statement).
    \end{tablenotes}
\end{threeparttable}
\end{table*}
\bsp	
\label{lastpage}
\end{document}